\documentclass[aps,prl,twocolumn,superscriptaddress,amsmath,amssymb,floatfix]{revtex4-2}
\usepackage{graphicx}
\usepackage{bm}
\usepackage{xcolor}
\usepackage[colorlinks=true,linkcolor=blue,citecolor=blue,urlcolor=blue,bookmarksnumbered=true,bookmarksopen=true]{hyperref}

\newcommand{\tr}{\operatorname{tr}}

\begin{document}

\title{A graph-theoretic Chern rule for gapped twin singular flat bands}

\author{Xiaoting Zhou}
\email{x.zhou@northeastern.edu}
\affiliation{Department of Physics, Northeastern University, Boston, Massachusetts 02115, USA}
\affiliation{Quantum Materials and Sensing Institute, Northeastern University, Burlington, Massachusetts 01803, USA}
\author{Arun Bansil}
\email{ar.bansil@northeastern.edu}
\affiliation{Department of Physics, Northeastern University, Boston, Massachusetts 02115, USA}
\affiliation{Quantum Materials and Sensing Institute, Northeastern University, Burlington, Massachusetts 01803, USA}

\date{\today}

\begin{abstract}
Nearly flat Chern bands, which can host fractional Chern insulators, may be obtained by gapping the band touchings of singular flat bands, but their Chern numbers are usually determined numerically. Line graphs of subdivided bipartite lattices host twin singular flat bands. On the star lattice and its square analog we prove that the Chern number of a gapped twin is the clique chirality it selects, which follows directly from three real-space features defined for any regular root graph. Tuning a hopping ratio through a triple point reverses the Chern number of one twin only, at fixed time-reversal breaking, and switches one-way edge channels on and off. Our study thus turns band topology into a design strategy for molecular frameworks and synthetic lattices.
\end{abstract}

\maketitle

Flat bands quench the kinetic energy, so the quantum geometry of their Bloch states controls their physics. The quantum metric $g$ bounds the superfluid weight of a flat-band superconductor from below~\cite{Peotta2015,Torma2022}, and a flat Chern band behaves like a Landau level when its Berry curvature $\Omega$ saturates the trace inequality $\tr g\ge|\Omega|$~\cite{Roy2014,Claassen2015,Jackson2015,Wang2021}. With finite-range hopping, exact flat bands admit compact localized states (CLSs), eigenstates confined by destructive interference~\cite{Mielke1991a,Bergman2008,Rhim2019,Kollar2020,Calugaru2022}. For an isolated flat band the CLSs are complete and give a smooth Bloch gauge over the whole Brillouin zone (BZ), so its Chern number $C$ vanishes~\cite{Chen2014,Rhim2019}, unlike for a Landau level. A singular flat band escapes this constraint. Its Bloch states are singular where it touches a dispersive band, so its CLSs are incomplete~\cite{Rhim2019,Rhim2020}, and gapping the touching point can yield a Chern band~\cite{Rhim2019}.

In line graphs and subdivision graphs of bipartite lattices (called split graphs in Ref.~\cite{Ma2020}), inversion eigenvalues show that a symmetry-preserving spin-orbit coupling makes the gapped quasiflat band $\mathbb{Z}_2$ nontrivial~\cite{Ma2020}. Two questions remain. The first is which Chern number a given perturbation produces. Inversion eigenvalues fix $C$ modulo~2, gapping a quadratic band touching point transfers one unit of Chern number~\cite{Sun2009}, and symmetry indicators turn the rotation eigenvalues of an isolated band into $C$ modulo the order of the rotation~\cite{Fang2012}, with topological quantum chemistry tying these eigenvalues to orbitals, including on line graphs~\cite{Bradlyn2017,Chiu2020}. None of these tools, however, identifies the eigenvalues acquired by a singular band, which are fixed only once the touching is gapped and in practice are found by numerical band calculations. The second question is how flat, and how close to a Landau level, the resulting Chern band can be.

\begin{figure}[t]
\includegraphics[width=\columnwidth]{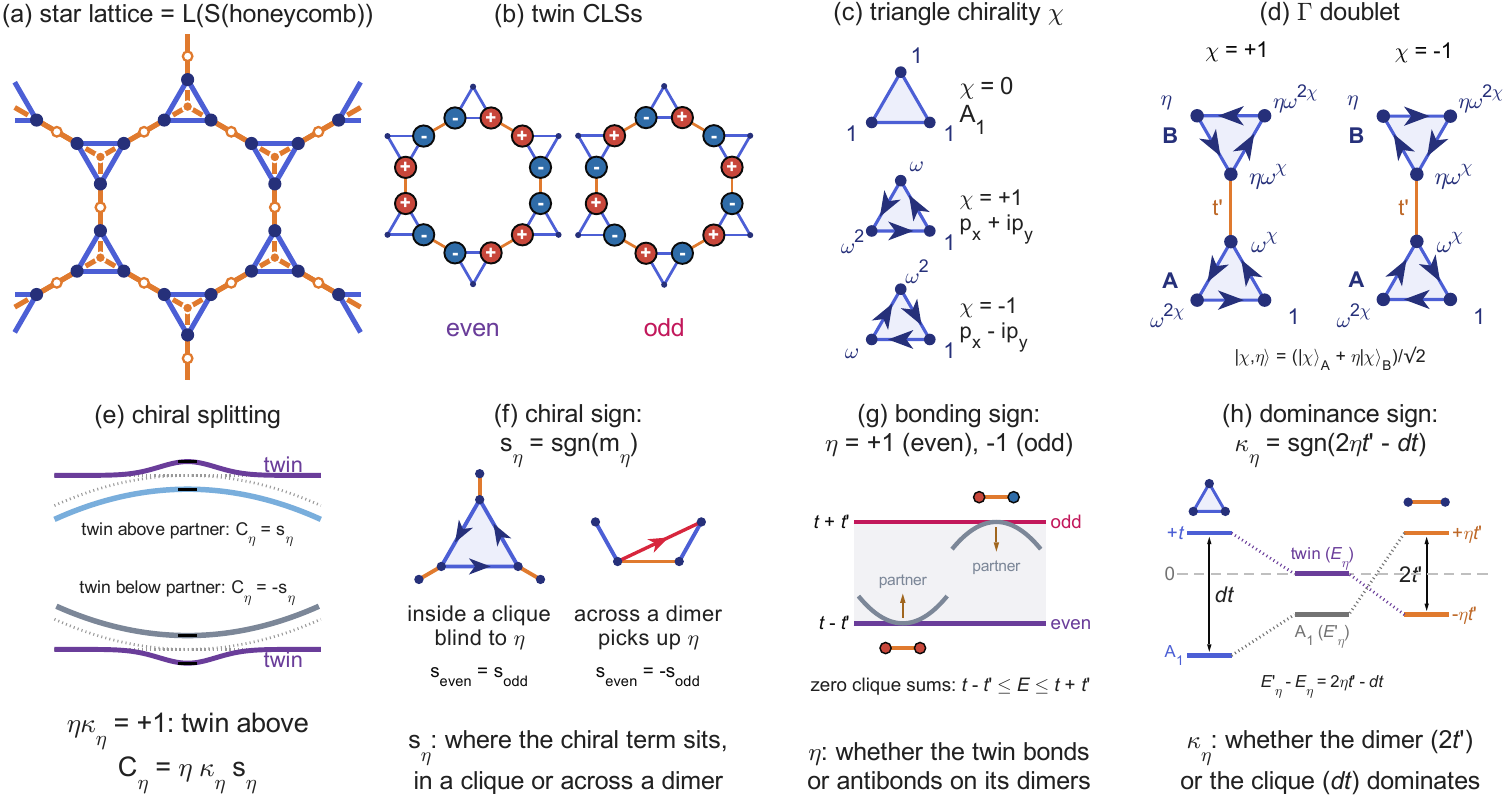}
\caption{\label{fig:1}Real-space origin of the Chern rule. (a)~Star lattice $L(S(G))$ of the honeycomb root $G$ (dashed), with cliques $K_3$ (indigo) joined by dimers (orange). (b)~CLSs of the even (left) and odd (right) twins on one dodecagon, with amplitudes $+1$ (red) and $-1$ (blue), symmetric ($\eta=+1$) or antisymmetric ($\eta=-1$) across every dimer, which defines the dimer parity $\eta$. (c)~Triangle orbitals, the nonchiral $A_1$ and the chiral $\chi=\pm1$ circulating in opposite senses (arrows). (d)~The $\Gamma$ doublet $|\chi=\pm1,\eta\rangle$, with the same chirality $\chi$ on both triangles and dimer partners differing by $\eta$. (e)~A chiral term splits the doublet, shifting the chirality $s_\eta$ up and $-s_\eta$ down. The twin selects the member on its own side, $s_\eta$ if it lies above its partner (top) and $-s_\eta$ if below (bottom), and this chirality is its Chern number. (f)--(h)~The three signs of $C_\eta=\eta\kappa_\eta s_\eta$, each read in real space. (f)~Chiral sign: a loop current inside a clique is blind to $\eta$, whereas a next-nearest-neighbor hop across a dimer picks up $\eta$. (g)~Bonding sign: states with zero clique sums lie between the even twin at $t-t'$, bonding on every dimer, and the odd twin at $t+t'$, antibonding, so the dimers push the partner up from the even twin and down from the odd twin (arrows). (h)~Dominance sign: the $A_1$ level $E'_\eta$ lies below the twin at $E_\eta$, $\kappa_\eta=-1$, when the clique splitting $dt$ exceeds the dimer splitting $2\eta t'$ (even twin at $t'=t$ shown). Level repulsion from this level overturns the dimer push only when it lies above the twin, which happens for the even twin on the dimer-dominated side.}
\end{figure}

We answer both questions above for the line graphs of subdivided root graphs $G$, denoted $L(S(G))$, in which cliques of mutually connected sites are linked by dimers [Fig.~\ref{fig:1}(a)]. The honeycomb root gives the star lattice (SL)~\cite{Yao2007,Ruegg2010,Wen2010,CrastoDeLima2019,Chen2012,Zeng2025}. Each such lattice hosts two exact flat bands, which we call twins. Their CLSs occupy the same closed loops and are symmetric across every dimer for the even twin and antisymmetric for the odd twin. The odd twin always touches a dispersive band (its partner) and is therefore singular. The even twin does so when $G$ is bipartite, as is the case for the honeycomb lattice.

At the BZ center $\Gamma$, each twin is degenerate with its partner in two states that circulate in opposite senses around the triangles, like the $p_x\pm ip_y$ orbitals. A chiral term splits this pair, and the twin selects the member shifted to its own side of the partner. We prove that the Chern number of the twin is the chirality of this member. It is fixed by three straightforward features of the graph, namely, where the chiral term sits, how the twin bonds on its dimers, and whether the cliques or the dimers dominate. The odd twin always lies above its partner, whereas the even twin lies below it when the cliques dominate and above it when the dimers dominate, so tuning a hopping ratio alone reverses its Chern number at fixed time-reversal breaking. Our Chern rule thus provides a direct mapping from the graph structure to the eigenvalues that symmetry indicators require. Since this mapping carries over to other regular root graphs with rotation symmetry, band topology can be read off, and designed, in real space. The same touching pins the flatness ratio near 1, which explains why such topological flat bands are only quasiflat.

\textit{Lattices from graphs.}---Placing a new vertex at the midpoint of every edge of a root graph $G$ gives its subdivision $S(G)$. Its line graph $L(S(G))$ has one site for each edge of $S(G)$, that is, for each half-edge of $G$, and joins two sites when their edges share a vertex [Fig.~\ref{fig:1}(a)]. The $d$ sites around a vertex of degree $d$ are all mutually joined and form a clique $K_d$, and the two sites on the halves of an edge are joined into a dimer. Every site thus lies in exactly one clique and one dimer. For the honeycomb root ($d=3$) the cliques are triangles, and the SL is realized in metal-organic frameworks~\cite{Zheng2007,Hu2023}. With hopping $-t$ on the clique bonds and $-t'$ on the dimer bonds ($t,t'>0$ unless stated otherwise),
\begin{equation}
H_0=-t\!\sum_{\langle ij\rangle\in\text{cliques}}\!\big(c_i^\dagger c_j+\mathrm{H.c.}\big)-t'\!\!\sum_{\langle ij\rangle\in\text{dimers}}\!\!\big(c_i^\dagger c_j+\mathrm{H.c.}\big). \label{eq:H0}
\end{equation}
By \emph{graph structure} we mean the connectivity together with these two weights.

The flat bands become visible once $H_0$ is written clique by clique and dimer by dimer. A clique $v$ contributes $tI-t\,\mathbf b_v\mathbf b_v^\dagger$, where $\mathbf b_v=(1,\dots,1)^T$ is its in-phase, or $A_1$, orbital and $\mathbf b_v^\dagger|u\rangle=\sum_{i\in v}u_i$ is the clique sum of a state $|u\rangle$. A dimer $e$ likewise contributes $t'I-t'\mathbf b_e\mathbf b_e^\dagger$ with its bonding orbital $\mathbf b_e=(1,1)^T$. Since every site lies in one clique and one dimer~\cite{Krausz1943,Harary1969}, the diagonal terms add up to $t+t'$ and
\begin{equation}
H_0=(t+t')\,I-t\,B_{\mathcal V}^\dagger B_{\mathcal V}-t'\,B_{\mathcal E}^\dagger B_{\mathcal E}, \label{eq:fact}
\end{equation}
where $B_{\mathcal V}=(\mathbf b_{v_1},\mathbf b_{v_2},\dots)^\dagger$ runs over all cliques, $B_{\mathcal E}=(\mathbf b_{e_1},\mathbf b_{e_2},\dots)^\dagger$ over all dimers, and $B=\binom{B_{\mathcal V}}{B_{\mathcal E}}$ is the incidence matrix of $S(G)$. This factorization is known for line graphs~\cite{Mielke1991a,Katsura2010,Mizoguchi2019} and holds at momentum $\mathbf k$ with the Bloch form $B(\mathbf k)$, as shown in the Supplemental Material (SM)~\cite[S1A]{SM}.

\begin{figure}[t]
\includegraphics[width=\columnwidth]{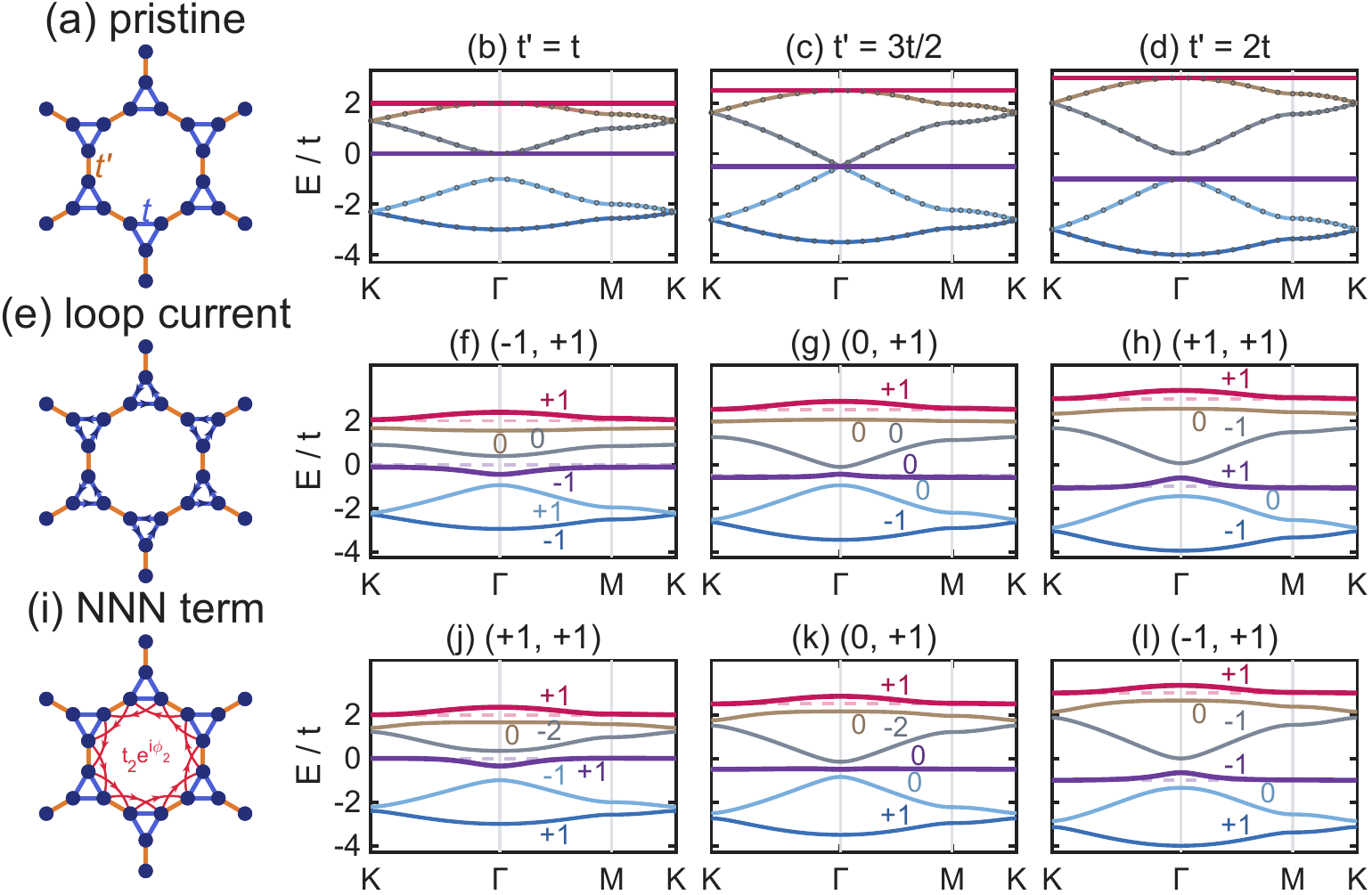}
\caption{\label{fig:2}Twin bands and Chern numbers. (a), (e), (i)~The pristine SL, the loop current $te^{i\phi_1}$ with $\phi_1=7\pi/90$, and the NNN term $t_2e^{i\phi_2}$ with $t_2=0.1t$ and $\phi_2=\pi/2$. Arrows point along the phase $e^{+i\phi}$. (b)--(d)~Pristine bands at $t'=t$ (star-lattice type), $3t/2$ (triple point), and $2t$ (double-kagome type), with Eq.~\eqref{eq:closed} as circles. Near $\Gamma$ the even twin (purple) lies below its partner at $t'=t$ and above it at $2t$, as given by Eq.~\eqref{eq:side}~\cite[S3]{SM}. (f)--(h), (j)--(l)~Bands with the loop current ($|m_\eta|=0.42t$) and the NNN term ($|m_\eta|=0.35t$), labeled by their Chern numbers, with $(C_{\rm even},C_{\rm odd})$ in the titles. Dashed lines mark the twins of the pristine SL.}
\end{figure}

The hoppings thus act only through the $A_1$ and bonding orbitals. A state with zero sum on every clique and on every dimer ($u_j=-u_i$ across each dimer) satisfies $B|u\rangle=\mathbf 0$. It is invisible to the hoppings and is an eigenstate at $t+t'$ for every momentum, so these states form the odd twin. For a $d$-regular root graph, each band $\varepsilon(\mathbf k)$ of the adjacency matrix of $G$ gives one pair of dispersive bands of $L(S(G))$ at $E=t+t'-\lambda$, with
\begin{equation}
\lambda^2-(dt+2t')\lambda+tt'(d-\varepsilon)=0
\label{eq:closed}
\end{equation}
\cite[S1B]{SM}. For the SL, $\varepsilon(\mathbf k)$ runs over the two honeycomb bands, and Eq.~\eqref{eq:closed} reproduces all dispersive bands [Figs.~\ref{fig:2}(b)--\ref{fig:2}(d)]. Since both hopping terms in Eq.~\eqref{eq:fact} can only lower the energy, no band lies above the odd twin.

\textit{Twins and partners.}---Keeping every clique sum zero but making every dimer bonding, $u_j=u_i$, gives a second flat band, the even twin. The cliques again contribute $t$, and each dimer now contributes $-t'$ instead of $+t'$, at every momentum. The two twins thus share the same clique pattern and differ only in their dimer parity $\eta=\pm1$, which places them at $E_\eta=t-\eta t'$.

Both twins have CLSs on the rings of $L(S(G))$ along which cliques and dimers alternate. These rings come from loops of $G$, and on the SL the hexagons of the honeycomb become dodecagons [Fig.~\ref{fig:1}(b)]. On each clique of the ring the state has amplitudes $+1$ and $-1$ on the two ring sites, so the clique sum vanishes and no amplitude leaks off the ring. Since every clique flips the sign and every dimer multiplies it by $\eta$, the odd ring closes on any loop and the even ring only on loops of even length. With $N_v$ vertices and $N_e$ edges of $G$ per cell, each twin consists of $N_e-N_v$ flat bands, which is positive for $d\ge3$ and one for the SL~\cite[S1C]{SM}.

A twin touches a dispersive band, its partner, where Eq.~\eqref{eq:closed} places a dispersive state at its energy. The uniform state of $G$, with $\varepsilon=d$ at $\Gamma$, gives $\lambda=0$, the energy of the odd twin. The staggered state, with opposite signs on neighboring vertices and $\varepsilon=-d$, exists if and only if $G$ is bipartite~\cite{Mohar1982} and gives $\lambda=2t'$, the energy of the even twin. For a non-bipartite root the even twin therefore has no partner and is trivial~\cite[S1C]{SM}. For the SL the staggered state also lies at $\Gamma$, so both twins touch their partners there. The ring CLSs then miss the non-contractible loop states, which wind around the lattice and enforce the touching~\cite{Bergman2008}, and the twins are singular~\cite{Rhim2019}~\cite[S6B]{SM}.

The odd twin lies above its partner, and the even twin lies below its partner at $t'=t$ and above it at $t'=2t$ [Figs.~\ref{fig:2}(b) and \ref{fig:2}(d)]. In between, at $t'=dt/2$, or $3t/2$ on the SL, Eq.~\eqref{eq:closed} has a double root at the energy of the even twin, so the even twin touches both bands at a triple point and changes its partner [Fig.~\ref{fig:2}(c)]~\cite[S2C]{SM}.

\textit{The touching point.}---At $\Gamma$ each twin is degenerate with its partner, and both states are built from the triangle orbitals. These are the $C_3$ eigenstates $|\chi\rangle=\sum_j\omega^{\chi j}|j\rangle/\sqrt3$ with eigenvalue $\omega^{-\chi}$, where $\omega=e^{2\pi i/3}$, $j=1,2,3$ runs counterclockwise, and $\chi=0,\pm1$ [Fig.~\ref{fig:1}(c)]~\cite[S2A]{SM}. The nonchiral orbital $A_1$ ($\chi=0$) lies at $-2t$, and the chiral orbitals $\chi=\pm1$, which circulate in opposite senses, are degenerate at $+t$. A twin sums to zero on every clique, so it carries no $A_1$ weight. At $\Gamma$, $C_3$ symmetry lets the dimers couple each $|\chi\rangle$ only to the orbital of the same chirality on the other triangle of the unit cell, giving $|\chi,\eta\rangle=(|\chi\rangle_A+\eta|\chi\rangle_B)/\sqrt2$ at $E_\eta=t-\eta t'$. For $\chi=\pm1$ these states form the degenerate doublet of the twin and its partner [Fig.~\ref{fig:1}(d)].

A chiral term that breaks time reversal splits the doublet, shifting the chirality $s_\eta$ up and $-s_\eta$ down [Fig.~\ref{fig:1}(e)]. The twin selects the member on its own side, $s_\eta$ if it lies above its partner and $-s_\eta$ if below. We show below that this selected chirality is the Chern number. The chiral term fixes $s_\eta$, so it remains to find on which side of its partner the twin lies.

By Eq.~\eqref{eq:fact}, a state with zero clique sums lies between $t-t'$ and $t+t'$, the energies of the even twin, bonding on every dimer, and of the odd twin, antibonding on every dimer. Away from the touching point, the Bloch phases on the dimers spoil the dimer parity of the partner and push it up from the even twin and down from the odd twin. Only clique sums can change this push. The partner acquires them only by mixing with $|A_1,-\eta\rangle$, the $A_1$ state from the same band of $G$ as the partner, whose uniform or staggered pattern gives it the dimer parity opposite to that of the twin~\cite[S3]{SM}. It lies at $E'_\eta=\eta t'-(d-1)t$. Relative to the twin, its clique sums lower it by $dt$ and its flipped dimers shift it by $2\eta t'$, so
\begin{equation}
\kappa_\eta=\mathrm{sgn}(E'_\eta-E_\eta)=\mathrm{sgn}(2\eta t'-dt)
\label{eq:kappa}
\end{equation}
tells on which side of the twin it lies. Level repulsion pushes the partner away from $|A_1,-\eta\rangle$. Expanding Eq.~\eqref{eq:closed} about the touching point gives both effects on the partner energy $E_{\rm p}$,
\begin{equation}
E_{\rm p}-E_\eta\simeq\frac{\eta\,t'\delta}{d}+\frac{2t'^2\delta}{d\,(E_\eta-E'_\eta)}=\frac{\eta\,tt'\delta}{E_\eta-E'_\eta},
\label{eq:side}
\end{equation}
where $\delta=d+\eta\varepsilon\ge0$ measures how far $\varepsilon$ has moved from its value $d$ in the uniform state or $-d$ in the staggered state~\cite[S3]{SM}. By Eq.~\eqref{eq:kappa}, $\mathrm{sgn}(E_{\rm p}-E_\eta)=-\eta\kappa_\eta$, so the twin lies above its partner for $\eta\kappa_\eta=+1$ and below it for $\eta\kappa_\eta=-1$. For the odd twin the cliques and the dimers both lower $|A_1,+\rangle$, so $\kappa_{\rm odd}=\mathrm{sgn}(-2t'-dt)=-1$ for all $t,t'>0$, the repulsion only weakens the dimer push, and the odd twin always lies above its partner. For the even twin, the cliques lower $|A_1,-\rangle$ by $dt$ while the dimers raise it by $2t'$. The even twin therefore lies below its partner when the cliques dominate, $dt>2t'$, and above it when the dimers dominate, because the repulsion from $|A_1,-\rangle$ above then wins over the dimer push. The two balance at the triple point $t'=dt/2$.

\textit{Chiral terms.}---We break time reversal at zero net flux with two Haldane-type chiral terms~\cite{Haldane1988}, a loop current inside every triangle and a phase on next-nearest-neighbor (NNN) bonds, each made of one triangle bond and one dimer bond [Figs.~\ref{fig:2}(e) and \ref{fig:2}(i)],
\begin{align}
H ={}& -t\!\sum_{\triangle,\triangledown}\!\sum_{\langle ij\rangle}\!\left(e^{i\phi_1}c_i^\dagger c_j+\mathrm{H.c.}\right)
 - t'\!\!\sum_{\langle ij\rangle_{\rm inter}}\!\!\left(c_i^\dagger c_j+\mathrm{H.c.}\right) \nonumber\\
 & - t_2\sum_{\langle\!\langle ij\rangle\!\rangle}\!\left(e^{i\phi_2}c_i^\dagger c_j+\mathrm{H.c.}\right), \label{eq:H}
\end{align}
with $j$ following $i$ counterclockwise around each triangle and each NNN path turning counterclockwise. We consider spinless particles with one rotation-invariant orbital per site and orient Chern numbers as in Ref.~\cite{Fukui2005}~\cite[S0]{SM}. Both terms keep the sixfold rotation symmetry of the SL, so at $\Gamma$ they shift the doublet members, which have different $C_3$ eigenvalues, by $\chi m_\eta$ with
\begin{equation}
m_\eta=\sqrt3\,(t\sin\phi_1-2\eta\,t_2\sin\phi_2) , \label{eq:mass}
\end{equation}
and $s_\eta=\mathrm{sgn}(m_\eta)$. The loop current stays inside a clique and cannot see the dimer parity, so it gives both twins masses of the same sign. An NNN hop crosses a dimer, where a twin contributes only the factor $\eta$. Since every triangle bond lies on two NNN paths of opposite sense, the NNN term acts on a twin as $-2\eta$ times a loop current and gives the two twins masses of opposite sign [Fig.~\ref{fig:1}(f)]~\cite[S4]{SM}.

\textit{Chern rule.}---The selected chirality is the Chern number of the twin under weak gapping, with $|m_\eta|$ small compared with the level spacings around it~\cite[S5A]{SM}. Three facts show it. First, the pristine $H_0$ is real, that is, symmetric under $C_2\mathcal T$, the product of twofold rotation and time reversal. Away from $\Gamma$ the twin therefore stays real to leading order and carries no Berry curvature, while the gapped quadratic touching transfers exactly one unit, $|C|=1$~\cite{Sun2009}. Second, the $C_6$ symmetry indicator~\cite{Fang2012} relates the Chern number of an isolated band to its rotation eigenvalues, $\xi_\Gamma\theta_K\zeta_M=e^{-i\pi C/3}$, with $\xi_\Gamma$ for $C_6$ at $\Gamma$, $\theta_K$ for $C_3$ at the BZ corner $K$, and $\zeta_M$ for $C_2$ at the edge center $M$~\cite[S5A]{SM}. Third, these eigenvalues of the twin are fixed in real space. A $C_6$ rotation about the dodecagon center exchanges the two triangles of a cell, so at $\Gamma$ the selected member has $\xi_\Gamma=\eta\,\omega^{\chi}$. At $K$ and $M$ the twin is a Bloch sum of its CLS, with $\theta_K=1$ and $\zeta_M=-\eta$ independent of $t'$ and of the chiral term. The product is $-\eta^2\omega^\chi=e^{-i\pi\chi/3}$, so the dimer parity cancels and $C\equiv\chi$ modulo 6. With $|C|=1$ and the selected chirality $\chi=\eta\kappa_\eta s_\eta$,
\begin{equation}
C_\eta=\eta\,\kappa_\eta\,s_\eta .
\label{eq:rule}
\end{equation}
The three factors are the bonding sign $\eta$, the dominance sign $\kappa_\eta$, and the chiral sign $s_\eta$ [Figs.~\ref{fig:1}(f)--\ref{fig:1}(h)]. The rule also holds for $C_6$-breaking deformations that do not close the gap~\cite[S5A]{SM}, and it carries over to the square-lattice analog~\cite[S5D]{SM}.

The weak-gapping condition becomes restrictive only near the triple point, where the spacing $|E'_{\rm even}-E_{\rm even}|=|2t'-dt|$ to the $A_1$ level closes. For $|2t'-dt|<|m_{\rm even}|$ this level lies between the two split members of the even doublet, and the even twin selects it instead. Then $\chi=0$ and the indicator gives $C_{\rm even}=0$~\cite[S5B]{SM}, so across the triple point the even twin passes from $C=-s_{\rm even}$ to $+s_{\rm even}$ through a trivial interval.

Figure~\ref{fig:2} confirms the rule. The loop current gives $s_{\rm even}=s_{\rm odd}=+1$ and hence $(C_{\rm even},C_{\rm odd})=(-1,+1)$ at $t'=t$ and $(+1,+1)$ at $t'=2t$, while the NNN term gives $s_{\rm even}=-s_{\rm odd}=-1$ and hence $(+1,+1)$ and $(-1,+1)$. At $t'=dt/2$ the even twin is trivial for both terms [Figs.~\ref{fig:2}(f)--\ref{fig:2}(h) and \ref{fig:2}(j)--\ref{fig:2}(l)]. With both terms present, the two masses of Eq.~\eqref{eq:mass} vanish on different lines, so the two Chern numbers can be set independently~\cite[S5C and Fig.~S5]{SM}.

\begin{figure}[t]
\includegraphics[width=\columnwidth]{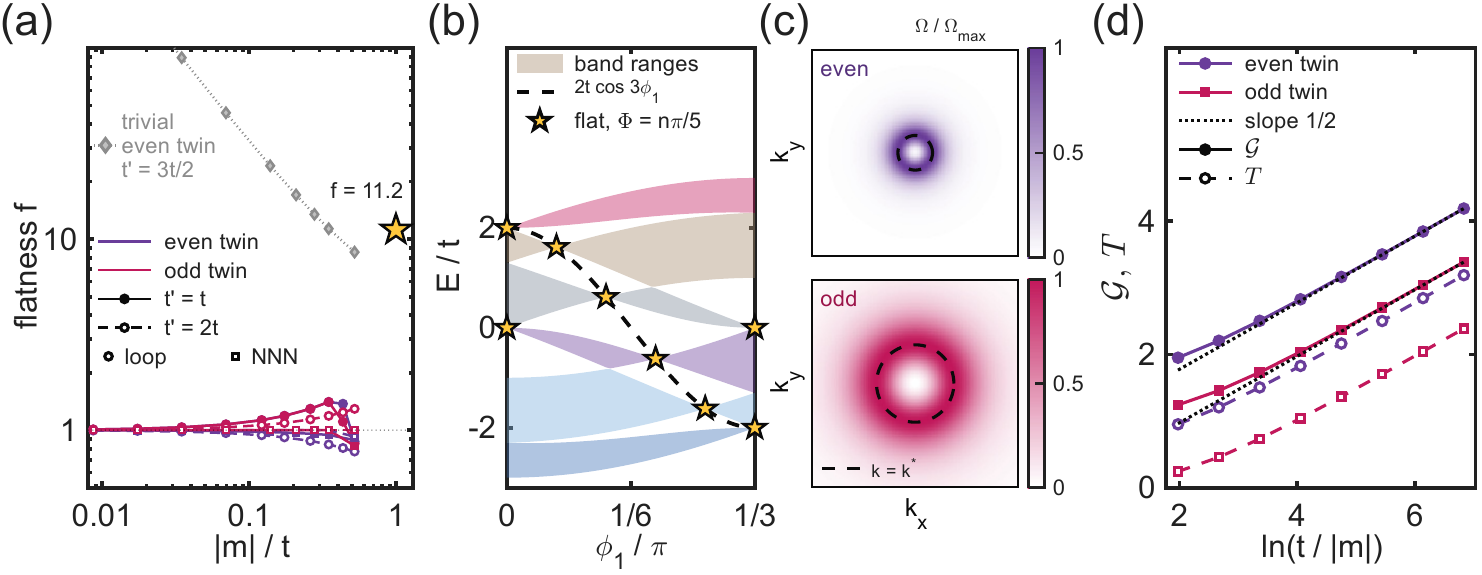}
\caption{\label{fig:3}Flatness and geometry of the gapped twins. (a)~Flatness ratio $f=\Delta_{\rm iso}/W$ of both twins versus $|m|$ for the loop current (circles) and the NNN term (squares) at $t'=t$ (filled) and $2t$ (open), with the trivial even twin at the triple point (gray diamonds) and a trivial loop band (star, $\phi_1=7\pi/36$). (b)~Loop-current bands versus $\phi_1$ at $t'=t$, with exact flat bands as stars~\cite[S6B]{SM}. (c)~Berry curvature of the twins (NNN term, $t'=t$, $|m|=0.017t$), with the ring $k=k^*$ dashed. (d)~$\mathcal G$ and $T$ versus $\ln(t/|m|)$.}
\end{figure}

\textit{Flatness and geometry.}---The same touching limits the flatness of any $C_2\mathcal T$-symmetric singular flat band with a twofold quadratic touching. The parent band is real, so a perturbation odd under $C_2\mathcal T$, such as a chiral term that preserves $C_2$, has no expectation value in it. To first order the band therefore stays flat away from the touching point while the doublet splits symmetrically. The bandwidth $W$ and the isolating gap $\Delta_{\rm iso}$ are then both $|m|$, and $f=\Delta_{\rm iso}/W=1+O(|m|/t)$ [Fig.~\ref{fig:3}(a)]~\cite[S6A]{SM}, as for the quasiflat bands of Ref.~\cite{Ma2020}. Strongly flat Chern bands require going beyond first order, as in models with tuned longer-range hopping~\cite{Tang2011,Sun2011,Neupert2011} or under circularly polarized light~\cite{Oka2009,Kitagawa2011}, whereas a band that stays exactly flat while isolated is trivial [Fig.~\ref{fig:3}(b)]~\cite[S6B]{SM}.

In the same two-band picture the Berry curvature is confined to a ring of radius $k^*\propto|m|^{1/2}$ that carries the full unit of flux, and outside it $\tr g\to1/k^2$~\cite[S7]{SM}. The integrated metric $\mathcal G=\int\tr g\,d^2k/2\pi$ therefore grows as $\tfrac12\ln(t/|m|)$ and the trace-condition violation $T=\int(\tr g-|\Omega|)\,d^2k/2\pi$ as $\mathcal G-1$ [Figs.~\ref{fig:3}(c) and \ref{fig:3}(d)]. The gapped twin therefore stays far from a Landau level, consistent with the logarithmic enhancement known for singular flat bands~\cite{Jiang2025}.

\textit{Discussion and Outlook.}---In line graphs of subdivided bipartite root graphs, the dimers give twins of opposite dimer parity, and a gapped twin selects the clique chirality shifted to its own side of the partner. Its Chern number is thus fixed in real space by three features of the graph, a rule that carries over to other regular root graphs with rotation symmetry~\cite[S5D]{SM}.

The Chern number becomes a design parameter. Its sign relative to the chiral term, $C_\eta/s_\eta=\eta\kappa_\eta$, is set by the graph and the real hoppings alone. In frameworks of molecular cores and linkers, the choice of linker sets this sign for one twin~\cite{Mazumdar2026}, and near the triple point, strain or pressure can reverse it, while the time-reversal breaking stays fixed~\cite{Sharma2025,Mukherjee2026}. Across the triple point, a single real coupling therefore switches one-way edge channels on and off~\cite[S8]{SM}, and in artificial lattices with tunable couplings these channels can be reconfigured. The reversal of the twin itself is seen by band-resolved probes. When the twin alone is populated, circular dichroism integrated over frequency measures its Chern number~\cite{Tran2017,Asteria2019} and should reverse across the triple point.

Metal-organic frameworks such as Ni$_3$(HITP)$_2$~\cite{Hu2023}, molecular crystals such as Mo$_3$S$_7$(dmit)$_3$~\cite{Jacko2015,Merino2018}, star-lattice magnets~\cite{Zheng2007,Sorolla2020,Ishikawa2024}, and optical~\cite{Jotzu2014,Aidelsburger2011} and photonic~\cite{Leykam2018} lattices offer platforms in which the graph, and with it the topology, can be engineered. The first-principles bands of Ni$_3$(HITP)$_2$ lie on the dimer-dominated side of the triple point~\cite{Hu2023}, so a loop current should give its two twins the same Chern number and an NNN term opposite ones. In Mo$_3$S$_7$(dmit)$_3$ the spin-conserving part of the spin molecular-orbit coupling acts as a loop current of opposite sign for the two spins~\cite{Jacko2015,Merino2018}, so the two spins carry opposite Chern numbers.

\let\origaddcontentsline\addcontentsline\renewcommand{\addcontentsline}[3]{}
\begin{acknowledgments}
We thank Baokai Wang for useful discussions. This work was supported by the National Science Foundation through Expand-QISE award OMA-2329067 and by the Massachusetts Technology Collaborative through award MTC-22032.
\end{acknowledgments}

\clearpage
\onecolumngrid
\let\addcontentsline\origaddcontentsline
\setcounter{secnumdepth}{3}
\setcounter{equation}{0}
\setcounter{figure}{0}
\setcounter{table}{0}
\renewcommand{\thefigure}{S\arabic{figure}}
\renewcommand{\theequation}{S\arabic{equation}}
\renewcommand{\thetable}{S\arabic{table}}
\renewcommand{\thesection}{S\arabic{section}}
\makeatletter\renewcommand{\p@subsection}{\thesection}\makeatother   

\begin{center}
{\large\bfseries Supplemental Material for ``A graph-theoretic Chern rule for gapped twin singular flat bands''\par}
\medskip
Xiaoting Zhou and Arun Bansil
\end{center}
\bigskip
\tableofcontents
\bigskip

This Supplemental Material gives the details of the results in the main text, in the same order. Section~S1 builds the lattices and their twin flat bands from the root graph. Section~S2 describes the two forms of the band structure considered and the triple point that separates them, where the even twin changes partner. Section~S3 derives the side of the partner from the graph structure, and Sec.~S4 the action of the chiral terms. Section~S5 proves the Chern rule and tests it, Sec.~S6 treats the limits of flatness, Sec.~S7 the quantum geometry of the gapped twins, and Sec.~S8 the edge channels that are switched on and off by a real coupling.

\setcounter{section}{-1}
\section{Conventions}\label{sm:sec:conv}
Hoppings enter as $-t$, $-t'$, $-t_2$ [Eq.~(6) of the main text]. $C_n$ denotes the active counterclockwise rotation by $2\pi/n$. For $C_6$ the center is the dodecagon center, and for $C_3$ the triangle center. Angular momentum is defined by $\hat R(\theta)=e^{-i\hat L_z\theta}$, so that a state with $\hat C_n$ eigenvalue $e^{-2\pi i\ell/n}$ has angular momentum $\ell$. The triangle orbitals
\begin{equation}
|\chi\rangle=\frac1{\sqrt3}\sum_{j=1}^3\omega^{\chi j}|j\rangle,\qquad \omega=e^{2\pi i/3},
\end{equation}
with vertices $j$ numbered counterclockwise, are: $|0\rangle=(1,1,1)/\sqrt3$, $|{+1}\rangle=(\omega,\omega^2,1)/\sqrt3$, and $|{-1}\rangle=(\omega^2,\omega,1)/\sqrt3$. Under $\hat C_3$ ($j\to j+1$) they have eigenvalues $\omega^{-\chi}$, i.e., $\ell=\chi$, and $|{+1}\rangle$ transforms like $p_x+ip_y$. Chern numbers are oriented as in the Fukui--Hatsugai--Suzuki method~\cite{Fukui2005}, $C=\int\Omega\,d^2k/2\pi$ with $\Omega=-i\,\tr(P[\partial_xP,\partial_yP])$ for a right-handed $(k_x,k_y)$, and $C^{(n)}$ denotes the Chern number of band $n$, counted from the bottom. A numeral subscript on $C$ denotes a rotation ($C_2$, $C_3$, $C_4$, $C_6$), and $C_\eta$ the Chern number of the twin of dimer parity $\eta$. The subscripts even and odd stand for $\eta=+1$ and $-1$ throughout ($\kappa_{\rm even}$, $m_{\rm even}$, $c_{\rm even}$, \dots). In these conventions the $C_6$ indicator of an isolated band reads $e^{-i\pi C/3}=\xi_\Gamma\theta_K\zeta_M$, with $\xi_\Gamma$, $\theta_K$, $\zeta_M$ the $C_6$, $C_3$, and $C_2$ eigenvalues at $\Gamma$, $K$, and $M$. The sign of the exponent is derived in Sec.~S5A (step 2), and Eq.~(33) of Ref.~\cite{Fang2012} has the opposite sign, which corresponds to the opposite orientation. Lengths are in units of the nearest-neighbor bond, so the lattice constant of the SL is $a=|\mathbf a_1|=2+\sqrt3$. Energies are in units of $t$ where no unit is given.

\section{Lattices from graphs and their twin flat bands}

\subsection{Subdivision, line graph, and factorization}\label{sm:sec:fact}
Let $G$ be a connected, periodic graph with $N_v$ vertices and $N_e$ edges per unit cell and vertex degrees $d_v$. Where a closed-form spectrum is needed, we take $G$ $d$-regular. Its subdivision $S(G)$ places a new vertex at the midpoint of every edge, so $S(G)$ is bipartite, with vertex classes $\mathcal V$ (the vertices of $G$) and $\mathcal E$ (the midpoints). The sites of the line graph $L(S(G))$ are the edges of $S(G)$, i.e., the half-edges $(v,e)$ of $G$ with $v\in e$, so there are $2N_e$ sites per cell. Two half-edges are adjacent in $L(S(G))$ when they share a vertex $v$, forming a clique, the complete graph $K_{d_v}$, or an edge $e$, forming a dimer. Every site therefore belongs to exactly one clique and one dimer. For $G$ = honeycomb, the cliques are the triangles of the star lattice (SL) and the dimers are its inter-triangle bonds. For $G$ = square and triangular, the cliques are $K_4$ and $K_6$, respectively.

We define the incidence matrices $(B_{\mathcal V})_{v,(v',e)}=\delta_{vv'}$ and $(B_{\mathcal E})_{e,(v,e')}=\delta_{ee'}$. Row $v$ of $B_{\mathcal V}$ is $\mathbf b_v^\dagger$, where $\mathbf b_v$ is 1 on the $d_v$ sites of the clique at $v$ and 0 elsewhere, and row $e$ of $B_{\mathcal E}$ is $\mathbf b_e^\dagger$, where $\mathbf b_e$ is 1 on the two sites of the dimer on $e$. Equivalently, $B_{\mathcal V}^\dagger=(\mathbf b_{v_1},\mathbf b_{v_2},\dots)$ and $B_{\mathcal E}^\dagger=(\mathbf b_{e_1},\mathbf b_{e_2},\dots)$, as in Eq.~(2) of the main text. Acting on a state, $B_{\mathcal V}|u\rangle$ lists the overlaps $\mathbf b_v^\dagger|u\rangle$, that is, the sums of the amplitudes over each clique, and $B_{\mathcal E}|u\rangle$ lists the sums over each dimer.

A clique of degree $d_v$ contributes the block $t\,\openone-t\,\mathbf b_v\mathbf b_v^\dagger$ on its sites, because every site of the clique is joined to every other. For a triangle,
\begin{equation}
-t\begin{pmatrix}0&1&1\\1&0&1\\1&1&0\end{pmatrix}=t\,\openone-t\,\mathbf b_v\mathbf b_v^\dagger,\qquad \mathbf b_v=(1,1,1)^T. \label{sm:eq:block}
\end{equation}
The second term acts only on the $A_1$ orbital $\mathbf b_v/\sqrt{d_v}$ and lowers it by $d_vt$, to $-(d_v-1)t$. The $d_v-1$ orbitals orthogonal to $\mathbf b_v$ stay at $+t$. A dimer contributes $t'\openone-t'\,\mathbf b_e\mathbf b_e^\dagger$, with its bonding orbital $\mathbf b_e/\sqrt2$ at $-t'$ and its antibonding orbital at $+t'$. Summing the blocks gives $\sum_v\mathbf b_v\mathbf b_v^\dagger=B_{\mathcal V}^\dagger B_{\mathcal V}=\openone+A_{\mathcal V}$ and $\sum_e\mathbf b_e\mathbf b_e^\dagger=B_{\mathcal E}^\dagger B_{\mathcal E}=\openone+A_{\mathcal E}$, where $A_{\mathcal V,\mathcal E}$ are the adjacency matrices within the cliques and dimers. The identity on the diagonal appears because each site lies in exactly one clique and one dimer. The Hamiltonian with clique hopping $t$ and dimer hopping $t'$, $H_0=-tA_{\mathcal V}-t'A_{\mathcal E}$, is therefore
\begin{equation}
H_0=(t+t')\openone-B^\dagger\mathcal WB,\qquad B=\begin{pmatrix}B_{\mathcal V}\\B_{\mathcal E}\end{pmatrix},\quad \mathcal W=\begin{pmatrix}t\,\openone_{\mathcal V}&0\\0&t'\,\openone_{\mathcal E}\end{pmatrix}, \label{sm:eq:fact}
\end{equation}
which is Eq.~(2) of the main text, where the matrix of Eq.~(1) is written as a sum of clique and dimer blocks~\cite{Mielke1991a,Katsura2010,Mizoguchi2019}. The constant $t+t'$ is the same on every site only because every site has one clique and one dimer. If site $i$ belonged to $n_i$ cliques, its diagonal entry would be $n_it$, and a nonuniform $n_i$ would act as an on-site potential, so $H_0$ would no longer be a constant minus $B^\dagger\mathcal WB$.

For any state, Eq.~\eqref{sm:eq:fact} gives
\begin{equation}
\langle u|H_0|u\rangle=t+t'-t\sum_v\big|\mathbf b_v^\dagger u\big|^2-t'\sum_e\big|\mathbf b_e^\dagger u\big|^2 . \label{sm:eq:energy}
\end{equation}
For $t,t'>0$, the two sums can only lower the energy, so $H_0\le(t+t')\openone$. A state with $B|u\rangle=\mathbf 0$ satisfies $H_0|u\rangle=(t+t')|u\rangle$ and sits at the top of the spectrum. The same holds in the Bloch form, with $B\to B(\mathbf k)$, where $B$ is the unsigned incidence matrix of $S(G)$. Since every dimer has two sites, $B_{\mathcal E}^\dagger B_{\mathcal E}=2\Pi$, where $\Pi$ projects onto states that are symmetric across every dimer. For a normalized state with zero clique sums, Eq.~\eqref{sm:eq:energy} therefore gives $t+t'-2t'\langle u|\Pi|u\rangle$, which lies between $t-t'$ and $t+t'$. This is the window used in the main text to find the side of the partner.

On a dimer, the bonding and antibonding orbitals add up to the identity, $\mathbf b_e\mathbf b_e^\dagger+\tilde{\mathbf b}_e\tilde{\mathbf b}_e^\dagger=2\,\openone$ with $\tilde{\mathbf b}_e=(1,-1)^\dagger$. The dimer block can therefore also be written as $-t'\openone+t'\,\tilde{\mathbf b}_e\tilde{\mathbf b}_e^\dagger$, and we obtain
\begin{equation}
H_0=(t-t')\openone-t\,B_{\mathcal V}^\dagger B_{\mathcal V}+t'\,\tilde B_{\mathcal E}^\dagger\tilde B_{\mathcal E},\qquad \tilde B_{\mathcal E}^\dagger\tilde B_{\mathcal E}=2(\openone-\Pi). \label{sm:eq:fact2}
\end{equation}
A state with $B_{\mathcal V}|u\rangle=\tilde B_{\mathcal E}|u\rangle=\mathbf 0$ sits at exactly $t-t'$. The two factorizations, Eqs.~\eqref{sm:eq:fact} and \eqref{sm:eq:fact2}, give the odd and even twins of Sec.~S1C.

\subsection{Closed-form spectrum}\label{sm:sec:closed}
For $d$-regular $G$, the nonzero eigenvalues of $B^\dagger\mathcal WB$ coincide with those of $\mathcal W^{1/2}BB^\dagger\mathcal W^{1/2}$, which acts on the $N_v+N_e$ vertices of $S(G)$:
\begin{equation}
\mathcal W^{1/2}BB^\dagger\mathcal W^{1/2}=\begin{pmatrix}dt\,\openone&\sqrt{tt'}\,M\\ \sqrt{tt'}\,M^\dagger&2t'\,\openone\end{pmatrix},\qquad M=B_{\mathcal V}B_{\mathcal E}^\dagger . \label{sm:eq:BBt}
\end{equation}
Here $M$ is the unsigned $N_v\times N_e$ incidence matrix of $G$, and $MM^\dagger=d+A_G(\mathbf k)$. The block structure has a simple interpretation: a clique-like level at $\lambda=dt$ and a dimer-like level at $\lambda=2t'$, coupled by clique-dimer mixing of squared magnitude $tt'(d+\varepsilon)$ for each eigenvalue $\varepsilon(\mathbf k)$ of $A_G(\mathbf k)$. The Schur complement gives two eigenvalues $\lambda_\pm$ from $(\lambda-dt)(\lambda-2t')=tt'(d+\varepsilon)$:
\begin{equation}
\lambda_\pm=\frac{dt+2t'\pm\sqrt{(dt+2t')^2-4tt'(d-\varepsilon)}}{2},\qquad E=t+t'-\lambda , \label{sm:eq:closed}
\end{equation}
which are the roots of Eq.~(3) of the main text. In addition, there are $N_e-N_v$ eigenvalues $\lambda=2t'$, from the vectors $(0,y)$ with $My=0$, and the remaining $N_e-N_v$ states of $L(S(G))$ lie in $\ker B$ and have $\lambda=0$. This exhausts all $2N_e$ bands. For the honeycomb, $\varepsilon=\pm|\gamma(\mathbf k)|$ with $\gamma=1+e^{i\mathbf k\cdot(\mathbf a_1-\mathbf a_2)}+e^{-i\mathbf k\cdot\mathbf a_2}$. Equation~\eqref{sm:eq:closed} reproduces the numerical bands to $5\times10^{-15}t$ for $t'/t=0.3$, $1$, and $2.5$ (200 random $\mathbf k$). For comparison, the kagome lattice is $L(G)$ with $H_{\rm kag}=-tA_{L(G)}=2t-tM^\dagger M$, so its bands are $E=-t(1+\varepsilon)$ together with the flat band $E=2t$ ($\ker M$).

For $t'=t$, Eq.~\eqref{sm:eq:closed} follows from the characteristic polynomial of clique-inserted graphs, $L(S(G))$ of a regular $G$~\cite{Zhang2009}. On the honeycomb root it contains the known SL bands~\cite{Ruegg2010}, including the case $t'\neq t$ written out in Refs.~\cite{Wen2010,Hu2023}. We have not found the weighted form for general $d$-regular $G$ in the literature.

\subsection{Twin flat bands}\label{sm:sec:twins}
\emph{Lemma (parity selection).} If $H_0x=Ex$ and $B_{\mathcal V}x=0$, then $x$ has definite parity on every dimer, and $E=t\pm t'$.

\emph{Proof.} A state whose amplitudes cancel on every clique, $B_{\mathcal V}x=0$, is invisible to the clique term, and Eq.~\eqref{sm:eq:fact} reduces to $t'B_{\mathcal E}^\dagger B_{\mathcal E}x=2t'\Pi x=(t+t'-E)x$. Hence $\Pi x=0$, giving $E=t+t'$, or $\Pi x=x$, giving $E=t-t'$. $\square$

\emph{Proposition.} Let $G$ be connected and periodic with $N_e>N_v$ per cell (average degree above two), with hopping $t$ on all cliques and $t'$ on all dimers. Then:
\begin{enumerate}
\item[(i)] Exact flat bands exist at $E_{\rm odd}=t+t'$ and $E_{\rm even}=t-t'$, each with multiplicity $N_e-N_v$ per cell, with dimer parity $\eta=-1$ and $+1$, respectively. Regularity of $G$ is not required.
\item[(ii)] The odd CLSs are the alternating cycles of $S(G)$, which are always even. The even CLSs are $x=B_{\mathcal E}^\dagger y$, where $y$ is a sign-alternating closed walk of even length in $G$, i.e., an even cycle or a pair of odd cycles joined by a path. Each independent loop of $G$ thus carries one twin of each parity.
\item[(iii)] For $d$-regular $G$, the odd twin touches a dispersive band at $\Gamma$, and the even twin touches a dispersive band if and only if $G$ is bipartite, at $\Gamma$ when the lattice translations preserve the two sublattices of $G$, and at a time-reversal-invariant momentum on the BZ boundary otherwise.
\item[(iv)] If $G$ is bipartite with sublattices $A$ and $B$, let $\Sigma=+1$ ($-1$) on the half-edges $(v,e)$ with $v\in A$ ($v\in B$). Then $\Sigma H_0(t,t')\Sigma=H_0(t,-t')$ and $\psi_{\rm odd}=\Sigma\psi_{\rm even}$.
\end{enumerate}

\emph{Proof.}
\begin{enumerate}
\item[(i)--(ii)] Odd family: $x\in\ker B$ gives $H_0x=(t+t')x$, and $\dim\ker B(\mathbf k)=2N_e-\mathrm{rank}\,B(\mathbf k)=N_e-N_v$ at generic $\mathbf k$. Even family: take $x=B_{\mathcal E}^\dagger y$ with $My=0$. Then $B_{\mathcal V}x=My=0$ and $B_{\mathcal E}^\dagger B_{\mathcal E}x=B_{\mathcal E}^\dagger(B_{\mathcal E}B_{\mathcal E}^\dagger)y=2x$, so $H_0x=(t-t')x$, with multiplicity $\dim\ker M(\mathbf k)=N_e-N_v$. In real space, $\ker B$ and $\ker M$ are spanned by sign-alternating closed walks. Every cycle of $S(G)$ is even, while for $G$ the kernel of $M$ needs closed walks of even length.
\item[(iii)] A dispersive branch of Eq.~\eqref{sm:eq:closed} meets the odd band where $\lambda_-=0$, i.e., where $\varepsilon=d$. This happens at $\Gamma$, for the uniform state, for every connected $G$. A branch meets the even band where $\lambda_\pm=2t'$, which gives $tt'(-d-\varepsilon)=0$, i.e., $\varepsilon=-d$. For a connected periodic graph, $-d\in\mathrm{spec}\,A_G(\mathbf k)$ for some $\mathbf k$ if and only if $G$ is bipartite: the touching needs a staggered state, which exists only on a bipartite graph~\cite{Mohar1982}. It is a Bloch state at the momentum where the sublattice signs are periodic, $\Gamma$ for the honeycomb and $M$ for the square lattice. Otherwise $d+\varepsilon\ge d+\min\varepsilon>0$, the even band is isolated, and by Ref.~\cite{Chen2014} it has $C=0$.
\item[(iv)] Every clique lies in one sublattice, so $\Sigma A_{\mathcal V}\Sigma=A_{\mathcal V}$, and every dimer joins $A$ to $B$, so $\Sigma A_{\mathcal E}\Sigma=-A_{\mathcal E}$. Hence $\Sigma H_0(t,t')\Sigma=H_0(t,-t')$, and the odd CLS of $H_0(t,-t')$, at energy $t-t'$, is mapped onto the even CLS of $H_0(t,t')$. $\square$
\end{enumerate}
Because $\Sigma$ is diagonal and $\mathbf k$ independent, the twins of a bipartite $G$ have identical sublattice weights and quantum geometry, for every $t'$ and on both sides of the triple point.

\emph{Non-bipartite roots.} The even family keeps $\eta=+1$, since the Lemma leaves no other value. What changes is the touching: by (iii) of the Proposition, the even family is separated from all dispersive bands and is isolated, hence trivial~\cite{Chen2014}, and it stays trivial under any perturbation that does not close this gap. For the triangular root ($K_6$ cliques, $N_e-N_v=2$) at $t'=0.7t$ the even pair lies at $0.3t$, separated by $0.42t$ from the nearest dispersive band. With a clique loop current $\phi_1=0.05\pi$ on the hexagon perimeter of each $K_6$ (diagonals real), the even pair has total $C=0$, whereas the odd pair, which touches a dispersive band at $\Gamma$, becomes topological.

\emph{Twin map with chiral terms.} The loop current acts within the cliques and commutes with $\Sigma$. Every NNN bond joins the two ends of a dodecagon corner $a\to c\to b$ made of one triangle bond and one dimer bond, so its ends lie in opposite sublattices of $G$ and $\Sigma$ reverses it. Hence
\begin{equation}
\Sigma H(t,t',\phi_1,t_2)\Sigma=H(t,-t',\phi_1,-t_2), \label{sm:eq:Schiral}
\end{equation}
so the loop current is even under the twin map and the NNN term odd. This is the operator form of the parity identity of Sec.~S4.

\emph{Relationship to Ref.~\cite{Ma2020}.} Ma \emph{et al.} showed that line graphs of split (subdivision) graphs of bipartite lattices host flat bands at $t'=t$ that touch dispersive bands, and for the SL they showed with topological quantum chemistry and the inversion criterion that symmetry-preserving spin-orbit coupling makes both flat bands $\mathbb Z_2$ nontrivial. For $S_z$-conserving spin-orbit coupling, each spin sector is given by Eq.~(6) of the main text with small chiral phases, so their result implies odd Chern numbers of the twins at weak coupling, but the sign structure found here is invisible to that argument. The proposition adds the parity selection, the energies $t\pm t'$ for $t'\neq t$, the twin map $\Sigma$, and the extension to non-bipartite roots.

\subsection{Examples}
Table~\ref{sm:tab:graph} lists the twins of $L(S(G))$ for $G$ = honeycomb (the SL), square ($K_4$ cliques), and triangular ($K_6$ cliques) cases at $t'=0.7t$ on a $48^2$ $\mathbf k$-grid. In finite clusters, the odd and even CLSs of the SL and $L(S(\text{square}))$ have residuals $\|(H-E)\psi\|=0$, and $\|\Sigma\psi_{\rm even}-\psi_{\rm odd}\|=0$.
\begin{table}[h]
\caption{\label{sm:tab:graph}Twin flat bands of $L(S(G))$ at $t'=0.7t$, with their multiplicity per cell and their distance $\Delta$ to the nearest dispersive band.}
\begin{ruledtabular}
\begin{tabular}{lccccc}
$G$ & bipartite & $N_e-N_v$ & odd, $E=1.7t$ & even, $E=0.3t$ & even touching point \\ \hline
honeycomb & yes & 1 & $\Delta=0$ ($\Gamma$) & $\Delta=0$ ($\Gamma$) & $\Gamma$ \\
square & yes & 1 & $\Delta=0$ ($\Gamma$) & $\Delta=0$ ($M$) & $M$ \\
triangular & no & 2 & $\Delta=0$ ($\Gamma$) & $\Delta=0.42t$ & none (isolated) \\
\end{tabular}
\end{ruledtabular}
\end{table}

\section{Two forms of the band structure and the triple point}

\subsection{Six levels at $\Gamma$}\label{sm:sec:gamma}
At $\Gamma$, the Bloch phase is unity in every cell, so a $\Gamma$ state is fixed by its six amplitudes in one cell, three on each triangle $A$ and $B$. Label the sites of triangle $A$ as $A_1$, $A_2$, and $A_3$ (at $90^\circ,210^\circ,330^\circ$ about its center) and those of triangle $B$ as $B_1$, $B_2$, and $B_3$ (at $270^\circ,30^\circ,150^\circ$). Both are counterclockwise orderings, and $A_j$ and $B_j$ are dimer partners. On each triangle the amplitudes decompose into the orbitals $|\chi\rangle_{A,B}=\sum_j\omega^{\chi j}|A_j\rangle/\sqrt3$ (likewise for $B$). The intra-triangle hopping gives $-2t$ for $\chi=0$ and $+t$ for $\chi=\pm1$, and the dimer hopping has matrix elements ${}_A\langle\chi'|(-t'\sum_jc^\dagger_{A_j}c_{B_j})|\chi\rangle_B=-t'\delta_{\chi\chi'}$: it couples only equal chirality. $H_0(\Gamma)$ therefore splits into three $2\times2$ blocks and is diagonalized by
\begin{equation}
|\chi,\eta\rangle=\frac{|\chi\rangle_A+\eta|\chi\rangle_B}{\sqrt2} , \label{sm:eq:doublet}
\end{equation}
in which both triangles circulate in the same sense:
\begin{center}
\begin{tabular}{lccc}
state & energy & $t'=t$ & $C_6$ eigenvalue \\\hline
$|A_1,+\rangle$ & $-2t-t'$ & $-3t$ & $+1$\\
$|A_1,-\rangle$ & $-2t+t'$ & $-t$ & $-1$\\
$|{\pm1},+\rangle$ (even doublet) & $t-t'$ & $0,0$ & $\omega^{\pm1}$\\
$|{\pm1},-\rangle$ (odd doublet) & $t+t'$ & $2t,2t$ & $-\omega^{\pm1}$
\end{tabular}
\end{center}
These are the roots of Eq.~\eqref{sm:eq:closed} at $\varepsilon=\pm d$. A rotation by $\pi/3$ about the dodecagon center maps $A_j\to B_{j-1}$ and $B_j\to A_{j-1}$, so $\hat C_6|\chi,\eta\rangle=\eta\,\omega^{\chi}|\chi,\eta\rangle$, $\xi_\Gamma=\eta\,e^{2\pi i\chi/3}$. The six $\Gamma$ states carry the six distinct sixth roots of unity, so every $C_6$-symmetric Bloch Hamiltonian is diagonal in this basis at $\Gamma$. Each twofold level is the touching point of one twin with its partner. There the two cannot be distinguished, which is the immovable singularity of the flat band~\cite{Rhim2019}.

\subsection{Star-lattice and double-kagome type band structures}
For each band $\varepsilon(\mathbf k)$ of a bipartite $d$-regular root, Eq.~\eqref{sm:eq:closed} gives a lower branch from $\lambda_+$ and an upper branch from $\lambda_-$. Group the bands as the two lower branches with the even twin and the two upper branches with the odd twin. Both $\lambda_+$ and $\lambda_-$ are monotonic in $\varepsilon$, with $\lambda_+$ minimal and $\lambda_-$ maximal at $\varepsilon=-d$, where the roots are $2t'$ and $dt$. The top of the lower group is therefore $\max(t+t'-\max(2t',dt),\,t-t')$, and the bottom of the upper group is $t+t'-\min(2t',dt)$.
\begin{itemize}
\item For $2t'<dt$, both equal $t-t'$: the upper branch touches the even twin from above, and the six bands are connected. This is the familiar band structure of the SL at $t'=t$, which we call the star-lattice type.
\item For $2t'>dt$, the top of the lower group is the even twin at $t-t'$ and the bottom of the upper group is $t+t'-dt$. The two groups are separated by the gap
\begin{equation}
\Delta_{\rm groups}=2t'-dt . \label{sm:eq:groupgap}
\end{equation}
\end{itemize}
Numerically, for the SL ($d=3$, $\varepsilon$ sampled in steps of $10^{-3}$), the gap vanishes for $t'\le1.5t$ and equals $0.020t$, $0.200t$, and $1.000t$ at $t'=1.51t$, $1.6t$, and $2t$, respectively, as Eq.~\eqref{sm:eq:groupgap} predicts.

\begin{figure}[h]
\includegraphics[width=0.95\textwidth]{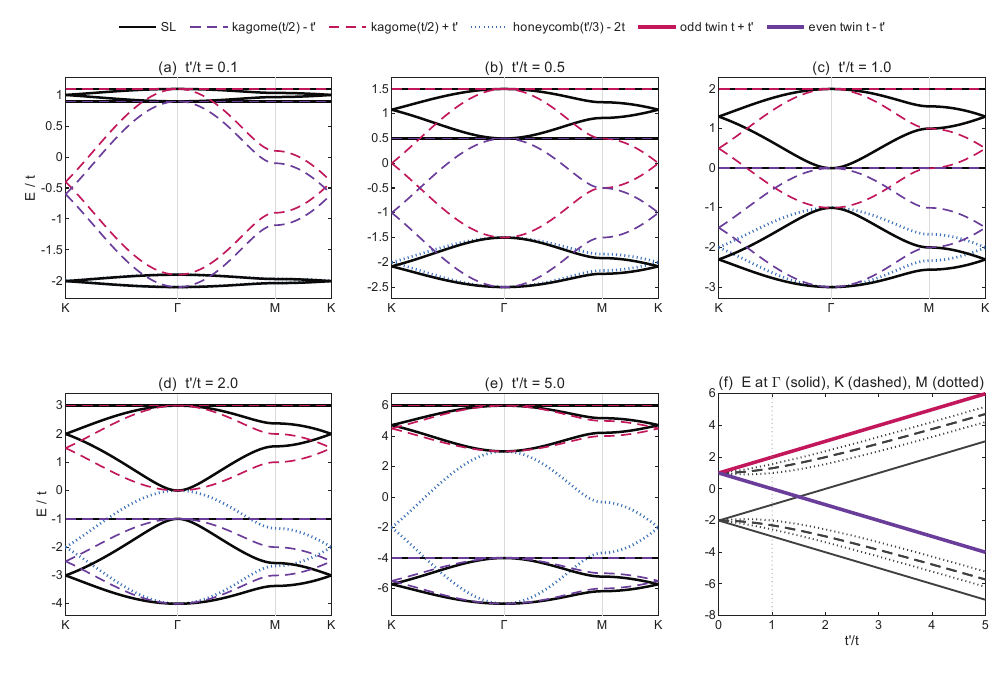}
\caption{\label{sm:fig:Skag}(a)--(e)~SL bands for $t'/t=0.1$, $0.5$, $1$, $2$, and $5$ (black solid lines). Two kagome copies with hopping $t/2$ at $-t'$ (purple dashed lines) and $+t'$ (raspberry dashed lines). Honeycomb bands of the triangle $A_1$ orbitals (dotted blue lines). (f)~Energies at $\Gamma$, $K$, and $M$ versus $t'/t$ from Eq.~\eqref{sm:eq:closed}, with the twins $t\pm t'$. Below the triple point ($t'<3t/2$) the band structure is of star-lattice type, and above it of double-kagome type.}
\end{figure}

Above the triple point, each group looks like a kagome band structure, and for $t'\gg t$ it becomes one exactly (Fig.~\ref{sm:fig:Skag}). Each dimer is then a super-site with a bonding orbital at $-t'$ ($\eta=+1$) and an antibonding orbital at $+t'$ ($\eta=-1$). Dimers that share a clique are coupled through the clique bond with amplitude $t/2$, so each sector is the line graph $L(G)$ of the root, a kagome lattice for the honeycomb, whose sites are the dimers and whose triangles are the cliques of the SL. Equation~\eqref{sm:eq:closed} gives $E\simeq\mp t'-\tfrac t2(1\pm\varepsilon)$ to $O(t^2/t')$. The bonding signs are uniform, and the antibonding signs can be gauged away because the product of the dimer signs around any kagome triangle is $+1$. The even (odd) twin is the flat band of the bonding (antibonding) copy, at $t-t'$ ($t+t'$). We call this a double-kagome type band structure.

\subsection{Triple point: the even twin changes partner}\label{sm:sec:triple}
At $\varepsilon=-d$, the touching momentum of the even twin, the roots of Eq.~\eqref{sm:eq:closed} are $\lambda=2t'$ (the twin) and $\lambda=dt$, i.e., the level $E'_{\rm even}=t'-(d-1)t$, which is the $A_1$ level $|A_1,-\rangle$ of odd dimer parity. For $t'<dt/2$, the branch $\lambda_-$ touches the even twin, and this dispersive band lies above it. For $t'>dt/2$, it is $\lambda_+$, and the band lies below. The even twin thus changes partner: the band above lets go of it and the band below takes over. At $\varepsilon=d$, the touching momentum of the odd twin, the roots are $0$ and $dt+2t'$, which never cross, so the partner of the odd twin is always the band directly below.

The handover passes through a threefold crossing. Near the touching point, $\varepsilon\simeq-d+\alpha k^2$ with $\alpha=a^2/4$ (Sec.~S3), and Eq.~\eqref{sm:eq:closed} becomes $(\lambda-dt)(\lambda-2t')=tt'\alpha k^2$. Away from the triple point, one root stays within $O(k^2)$ of $2t'$, the quadratic touching of the twin with its partner. At $2t'=dt$, the left-hand side becomes $(\lambda-dt)^2$, so
\begin{equation}
\lambda=dt\pm\sqrt{tt'\alpha}\,k ,
\end{equation}
and the two linear bands cross the flat twin at $E=t-t'$. The triple point is a pseudospin-1 crossing, a Dirac cone through a flat band. It is allowed as an accidental degeneracy because the crossing levels carry different $C_6$ eigenvalues, $-1$ and $\omega^{\pm1}$. The sequence is a quadratic touching with the partner above, a pseudospin-1 crossing, and a quadratic touching with the partner below [Figs.~2(b)--2(d) of the main text].

\section{Touching point and the side of the partner}\label{sm:sec:side}
\emph{Partner curvature.} Near $\Gamma$, the honeycomb dispersion is $|\gamma(\mathbf k)|=d-\alpha k^2+O(k^4)$, isotropic by $C_3$, with $\alpha=a^2/4=3.48205$. We write Eq.~\eqref{sm:eq:closed} as $p(\lambda,\varepsilon)\equiv\lambda^2-(dt+2t')\lambda+tt'(d-\varepsilon)=0$. The twin of parity $\eta$ is the root $\lambda_\eta=(1+\eta)t'$ at $\varepsilon=-\eta d$, which is $\Gamma$ for both twins on the honeycomb. The second root at the same point is $\lambda'_\eta=dt+(1-\eta)t'$, the level $E'_\eta=t+t'-\lambda'_\eta=\eta t'-(d-1)t$, which from Sec.~S2A is the $A_1$ level $|A_1,-\eta\rangle$ of opposite dimer parity. Along the partner branch $d\lambda/d\varepsilon=-\partial_\varepsilon p/\partial_\lambda p=tt'/(2\eta t'-dt)$, and near the touching point $\delta\varepsilon=\eta\alpha k^2$. Hence the partner disperses as
\begin{equation}
E_{\rm p}-E_\eta=c_\eta k^2,\qquad c_\eta=\frac{\eta\,tt'\alpha}{E_\eta-E'_\eta},\qquad E_\eta-E'_\eta=dt-2\eta t' . \label{sm:eq:cunified}
\end{equation}
For the odd twin $c_{\rm odd}=-tt'\alpha/(dt+2t')$, and for the even twin $c_{\rm even}=tt'\alpha/(dt-2t')$. At $t'=t$, $|c_{\rm even}/c_{\rm odd}|=(dt+2t')/(dt-2t')=5$. The numerical partner dispersion agrees with Eq.~\eqref{sm:eq:cunified} to $2\times10^{-4}$ (relative) in six directions for $t'/t=0.5$, $1$, $1.4$, $1.6$, and $2$, and $c_{\rm even}$ changes sign between $t'=1.4t$ ($+24.4$) and $1.6t$ ($-27.9$).

\emph{Side of the partner in two steps.} Equation~\eqref{sm:eq:cunified} fixes the side, and its two factors have a real-space origin. We write $\delta=d+\eta\varepsilon$, which vanishes at the touching point and equals $\alpha k^2$ near it. For a $d$-regular root
\begin{equation}
\delta=\frac{\sum_{e=(v,w)}|x_v+\eta x_w|^2}{\sum_v|x_v|^2}\ \ge0 .
\end{equation}
This is the Rayleigh quotient of the Laplacian ($\eta=-1$) or the signless Laplacian ($\eta=+1$) of $G$. It measures how far the state $\mathbf x$ of $G$ has moved from the uniform state ($x_w=x_v$, odd twin) or the staggered state ($x_w=-x_v$, even twin).
\begin{itemize}
\item \emph{Two forms of the energy.} Taking the expectation value of Eqs.~\eqref{sm:eq:fact} and \eqref{sm:eq:fact2} in any normalized state $u$,
\begin{equation}
E=t+t'-t\sum_v\Big|\sum_{i\in v}u_i\Big|^2-t'\sum_e|u_i+u_j|^2=t-t'-t\sum_v\Big|\sum_{i\in v}u_i\Big|^2+t'\sum_e|u_i-u_j|^2 . \label{sm:eq:twoforms}
\end{equation}
All terms of the first form vanish for the odd twin, and all terms of the second form vanish for the even twin. Over all states, the odd twin is the maximum of $H_0$, while the even twin is not an extremum, since clique sums lower the energy and dimer differences raise it. Equation~\eqref{sm:eq:twoforms} alone therefore fixes the side only for the odd twin. The two steps below treat both twins alike.
\item \emph{Step 1: zero clique sums.} For states with zero sum on every clique, Eq.~\eqref{sm:eq:twoforms} gives $t-t'\le E\le t+t'$. The even twin is the bottom and the odd twin the top of this set. A partner confined to it can only rise from the even twin and fall from the odd twin. This is the limit $t\to\infty$ of Eq.~\eqref{sm:eq:cunified}, in which the cliques are frozen at their upper level,
\begin{equation}
E_{\rm p}-E_\eta\;\to\;\frac{\eta\,t'\delta}{d} .
\end{equation}
\item \emph{Step 2: the only state with clique sums.} By Eq.~\eqref{sm:eq:BBt}, each dispersive state is built from a single eigenstate $\mathbf x$ of $A_G(\mathbf k)$: the $A_1$ orbitals $\mathbf b_v$ with weights $x_v$ and the bonding orbitals $\mathbf b_e$ with weights $x_v+x_w$. Different $\mathbf x$ do not mix. The only state with nonzero clique sums that the partner can mix with is thus the $A_1$ combination $\sum_vx_v\mathbf b_v/\sqrt d$ of its own $\mathbf x$. At the touching point, $\mathbf x$ is uniform for the odd twin and staggered for the even twin, so the $A_1$ amplitudes are equal or opposite at the two ends of every dimer. This state, $|A_1,-\eta\rangle$, has the opposite dimer parity $-\eta$ and lies at $E'_\eta=\eta t'-(d-1)t$. Relative to the twin, it turns on every clique sum and flips every dimer, so Eq.~\eqref{sm:eq:twoforms} gives $E'_\eta-E_\eta=-dt+2\eta t'$. On the SL the same selection follows from $C_3$: the $\mathbf k\cdot\mathbf p$ term is a current on the dimers, which flips the dimer parity, and a vector on each triangle, which changes $\chi$ by $\pm1$ (Sec.~S2A).
\item \emph{Sum of the two steps.} Expanding Eq.~\eqref{sm:eq:closed} to first order in $\delta$,
\begin{equation}
E_{\rm p}-E_\eta=\frac{\eta\,t'\delta}{d}+\frac{2t'^2\delta}{d\,(E_\eta-E'_\eta)}=\frac{\eta\,tt'\delta}{E_\eta-E'_\eta} , \label{sm:eq:twostep}
\end{equation}
which is Eq.~\eqref{sm:eq:cunified}. The second term is the repulsion from $|A_1,-\eta\rangle$, with the sign of $E_\eta-E'_\eta$. The numerator $dt$ of the sum never vanishes, so the side can change only where $|A_1,-\eta\rangle$ crosses the twin, at the triple point. For the odd twin, $E_\eta-E'_\eta=dt+2t'>2t'$, so the repulsion is always weaker than Step 1 and the partner always lies below. For the even twin, the repulsion adds to Step 1 for $dt>2t'$, and for $2t'>dt$ it points down and always wins, since then $|E_\eta-E'_\eta|<2t'$. The side found at the touching point holds over the whole BZ. Equation~\eqref{sm:eq:closed} gives exactly
\begin{equation}
E_{\rm p}-E_\eta=\frac{\eta\,tt'\delta}{\mu_\eta+\mathrm{sgn}(\mu_\eta)\sqrt{\mu_\eta^2+\eta\,tt'\delta}},\qquad \mu_\eta=\tfrac12(E_\eta-E'_\eta),
\end{equation}
whose denominator never vanishes, so the partner meets the twin only where $\delta=0$ and cannot cross it elsewhere.
\item \emph{Checks.} On the SL, the second-order $\mathbf k\cdot\mathbf p$ theory in the doublet reproduces both steps of Eq.~\eqref{sm:eq:twostep}. For $L(S(\text{square}))$, with $d=4$ and the even touching at $M$, Eq.~\eqref{sm:eq:twostep} agrees with the exact bands, and the even partner moves from above ($t'=t$) to below ($t'=3t$) as $2t'$ passes $dt=4t$.
\end{itemize}
In terms of signs, the twin lies above its partner when $c_\eta<0$ and below it when $c_\eta>0$. Since $tt'\alpha>0$,
\begin{equation}
-\mathrm{sgn}(c_\eta)=\eta\,\kappa_\eta,\qquad \kappa_\eta=\mathrm{sgn}(E'_\eta-E_\eta)=\mathrm{sgn}(2\eta t'-dt) , \label{sm:eq:sigmabk}
\end{equation}
so the twin lies above its partner for $\eta\kappa_\eta=+1$ and below it for $\eta\kappa_\eta=-1$. The bonding sign $\eta$ is the direction of Step 1, and the dominance sign $\kappa_\eta$ is the side of $|A_1,-\eta\rangle$ in Step 2. The derivation uses only Eq.~\eqref{sm:eq:closed}, so it holds for any connected $d$-regular root, with $\alpha$ the curvature of $\varepsilon(\mathbf k)$ at its extremum. For a bipartite root, the staggering $U$ obeys $A(\mathbf k+\mathbf Q)=-UA(\mathbf k)U$, so both twins see the same $\alpha$. Under $\Sigma$, $(\eta,t',t_2)\to(-\eta,-t',-t_2)$ leaves $\eta t'$, $\kappa_\eta$, and $m_\eta$ unchanged, so for $t'<0$, the result holds with $\eta$ replaced by $\mathrm{sgn}(\eta t')$, and the twins exchange roles.

\section{Chiral terms}\label{sm:sec:chiral}
\emph{Parity identity.} Let $\mathcal P_\eta$ be the projector onto states with zero amplitude sum on every triangle ($B_{\mathcal V}x=0$) and dimer parity $\eta$. Write the two chiral perturbations as $V_\triangle=\partial H/\partial\phi_1|_{\phi_1=0}$, which is $-it\nu$ on each triangle bond, and $V_{\rm N}=\partial H/\partial(t_2\sin\phi_2)$, which is $-i\nu_{\rm D}$ on each NNN bond, with $\nu$ and $\nu_{\rm D}$ the counterclockwise orientations around the triangle and the dodecagon. Then
\begin{equation}
\mathcal P_\eta V_{\rm N}\mathcal P_\eta=-\frac{2\eta}{t}\,\mathcal P_\eta V_\triangle\mathcal P_\eta,\qquad
\mathcal P_\eta A_{\rm N}\mathcal P_\eta=2\eta\,\mathcal P_\eta A_{\mathcal V}\mathcal P_\eta=-2\eta\,\mathcal P_\eta , \label{sm:eq:ruleop}
\end{equation}
where $A_{\rm N}$ is the real NNN adjacency. The physical picture behind the factor $-2\eta$ is simple. A loop current stays inside a clique, where a twin has no dimer to cross, so it cannot see $\eta$. An NNN hop crosses one dimer, and on a twin a dimer contributes only the factor $\eta$.

\emph{Proof.} Every NNN bond $a\to b$ is a dodecagon corner $a\to c\to b$ made of one triangle bond and one dimer bond. For $x,x'$ in the range of $\mathcal P_\eta$: if $c\to b$ is the dimer bond, then $x'_b=\eta x'_c$, and the bond contributes $\eta$ times the matrix element of the triangle bond $a\to c$. If $a\to c$ is the dimer bond, then $x_a^*=\eta x_c^*$, and the bond contributes $\eta$ times the matrix element of the triangle bond $c\to b$. Each triangle bond $i\to j$ therefore appears in exactly two NNN bonds, $i\to j'$ and $i'\to j$ (primes denote dimer partners), both on the dodecagon that borders $i\to j$. A counterclockwise traversal of the triangle is a clockwise traversal of this dodecagon, so $\nu_{\rm D}=-\nu$. Collecting the factors gives Eq.~\eqref{sm:eq:ruleop}. The second identity uses $A_{\mathcal V}x=-x$ for $B_{\mathcal V}x=0$, since $B_{\mathcal V}^\dagger B_{\mathcal V}=\openone+A_{\mathcal V}$. It reproduces the real shifts $E_{\rm odd,even}=t\pm t'\mp2t_2\cos\phi_2$. $\square$

\emph{Chiral terms are diagonal in $\chi$ at $\Gamma$.} Both chiral terms commute with $\hat C_6$, and the six states $|\chi,\eta\rangle$ carry six distinct $C_6$ eigenvalues (Sec.~S2A), so at $\Gamma$ the chiral terms shift each state without mixing it with its doublet partner or with the $A_1$ levels, to all orders. Inside a triangle, $\langle\chi|(-te^{i\phi_1\nu})|\chi\rangle=-2t\cos(\phi_1+2\pi\chi/3)=t\cos\phi_1+\sqrt3\,t\chi\sin\phi_1$ for $\chi=\pm1$, and $-2t\cos\phi_1$ for $\chi=0$. The loop current therefore splits the doublet exactly by $\pm\sqrt3\,t\sin\phi_1$ about a common shift $t\cos\phi_1$. By Eq.~\eqref{sm:eq:ruleop}, $\langle\eta,\chi|V_{\rm N}|\chi,\eta\rangle=-2\sqrt3\,\eta\chi$ per unit $t_2\sin\phi_2$. The shift of $|\chi,\eta\rangle$ is therefore $\chi m_\eta$ with
\begin{equation}
m_\eta=\sqrt3\left(t\sin\phi_1-2\eta\,t_2\sin\phi_2\right), \label{sm:eq:mass}
\end{equation}
which is Eq.~(7) of the main text, and the real parts $t\cos\phi_1$ and $t_2\cos\phi_2$ shift both members equally. The mass is independent of $t'$. Numerically, in the basis $|\chi,\eta\rangle$, at $t'=t$, $V_\triangle/\sqrt3t$ and $V_{\rm N}/\sqrt3$ are diagonal with entries $\chi$ and $-2\eta\chi$ on the doublets and zero on the $A_1$ levels. With the full Hamiltonian, the off-diagonal elements are below $3\times10^{-15}t$ for $\phi_1\le0.1\pi$, $t_2\le0.2t$, and both together. Projecting $V_{\rm N}$ and $V_\triangle$ onto the $\Gamma$ pair gives proportional blocks (residual $<3\times10^{-16}$) with ratio exactly $-2\eta$ for $t'/t=0.25$ to $10$.

\section{The Chern rule}

\subsection{Theorem and proof}\label{sm:sec:proof}
\emph{Theorem (Chern number of a gapped twin).} Let the chiral perturbation be weak enough that no gap closes except the quadratic touching at $\Gamma$ (in particular, $|m_\eta|$ is small compared with the level spacings around the touching point), and let the $A_1$ level lie outside the split doublet (otherwise, see Sec.~S5B). Let $\chi_{\rm sel}=\pm1$ be the triangle chirality of the $\Gamma$-doublet member that the twin selects when the touching point is gapped. Then
\begin{equation}
C_\eta=\chi_{\rm sel}=\eta\,\kappa_\eta\,s_\eta,\qquad s_\eta=\mathrm{sgn}(m_\eta), \label{sm:eq:thm}
\end{equation}
which is Eq.~(8) of the main text, with $\eta\kappa_\eta$ from Eq.~\eqref{sm:eq:sigmabk}. For $t,t'>0$, $\eta\kappa_\eta=+1$ for the odd twin and $\mathrm{sgn}(t'-3t/2)$ for the even twin.

\emph{Proof.}
\begin{enumerate}
\item \emph{$|C_\eta|=1$.} Write the chiral term as $\lambda V$, so that $m_\eta\propto\lambda$, and let $\ell_m=(2|m_\eta|/|c_\eta|)^{1/2}$ be the momentum scale of the gapped touching. $C_\eta$ does not change while the gap stays open, so it can be evaluated as $\lambda\to0$. We split the BZ at $|\mathbf k|=R\ell_m$, with $R\gg1$ fixed.
\begin{enumerate}
\item[(a)] \emph{Inner disk.} Near $\Gamma$ the twin and its partner are described by the two-band model of Eq.~\eqref{sm:eq:Heff} (Sec.~S7), in which the $\chi$-changing term has magnitude $\tfrac12|c_\eta|k^2$ (one band flat) and winds twice. After the rescaling $\mathbf k=\ell_m\boldsymbol\kappa$, $H-E_\eta\to(H-E_\eta)/|m_\eta|$, the corrections to $H_{\rm eff}$ vanish as $\lambda\to0$: these are the higher powers of $k$ and the coupling to the other bands, which stay at a finite distance away from the triple point. The flux of the twin in $|\mathbf k|<R\ell_m$ therefore tends to that of $H_{\rm eff}$, $\pm[1-(1+R^4)^{-1/2}]$.
\item[(b)] \emph{Outside.} For $|\mathbf k|\ge R\ell_m$ the twin is separated from its partner by at least $\tfrac12|c_\eta|k^2\gg|m_\eta|$, and from the other bands by a finite gap, so its projector is analytic in $\lambda$ there. At $\lambda=0$ the projector is real ($C_2\mathcal T$), and $\Omega=-i\,\tr(P[\partial_xP,\partial_yP])$ vanishes for a real projector. To first order, $|\Omega|\lesssim|m_\eta|/(|c_\eta|k^4)$, so the flux outside is $O(R^{-2})$, independent of $\lambda$.
\item[(c)] \emph{Order of limits.} Hence $C_\eta=\pm[1-(1+R^4)^{-1/2}]+O(R^{-2})+o(1)$, where $o(1)$ vanishes as $\lambda\to0$ at fixed $R$. Taking $\lambda\to0$ first, which leaves the integer $C_\eta$ unchanged, and then $R\to\infty$ gives $|C_\eta|=1$. The order matters: at fixed $\lambda$ the disk cannot be enlarged beyond the BZ, and only a disk that scales with $\ell_m$ captures the full flux.
\end{enumerate}
Numerically (NNN term, $t'=t$), the flux of the even twin in $|\mathbf k|<4\ell_m$ is $0.966$, $0.944$, and $0.939$ at $|m|=0.069t$, $0.017t$, and $0.0043t$, respectively, against the two-band value $0.938$. The odd twin behaves the same way, but its small $|c_{\rm odd}|$ makes $\ell_m$ large, and its flux in $|\mathbf k|<8\ell_m$ approaches the two-band value $0.984$ only at $|m|=0.0043t$ ($0.994$).
\item \emph{Symmetry indicator and its sign.} For $C_n$-symmetric bands, the rotation eigenvalues at the rotation-invariant momenta fix $C$ modulo $n$ through $e^{2\pi i\nu C/n}=\prod\lambda$~\cite{Fang2012}, where $\nu=\pm1$ depends on conventions. For $C_6$ the product is $\xi_\Gamma\theta_K\zeta_M$ (called $\eta_i$, $\theta_i$, $\zeta_i$ in Ref.~\cite{Fang2012}, renamed here to avoid a clash with the dimer parity). Our rotations are active and counterclockwise about the dodecagon center, $\hat C_nH(\mathbf k)\hat C_n^{-1}=H(R\mathbf k)$, and $C$ has the orientation of Sec.~S0. A band inversion at $\Gamma$ fixes $\nu$. Let two states $a$ and $b$ at $\Gamma$ have eigenvalues $\lambda_a$ and $\lambda_b$, and let $H=\begin{pmatrix}m&h_{ab}\\h_{ab}^*&-m\end{pmatrix}$ near $\Gamma$, with the other bands unchanged. Symmetry requires $h_{ab}(R\mathbf k)=(\lambda_a/\lambda_b)\,h_{ab}(\mathbf k)$, so $h_{ab}\propto k_-^p$ with $k_\pm=k_x\pm ik_y$ gives $\lambda_b/\lambda_a=e^{2\pi ip/n}$. In our orientation, the lower band of this model has $C=-\tfrac p2\,\mathrm{sgn}(m)$. When $m$ changes from negative to positive, the lower band exchanges $a$ for $b$ at $\Gamma$, its Chern number changes by $\Delta C=-p$, and its eigenvalue product changes by $\lambda_b/\lambda_a=e^{-2\pi i\Delta C/n}$. Hence $\nu=-1$:
\begin{equation}
e^{-2\pi iC/n}=\prod\lambda,\qquad e^{-i\pi C/3}=\xi_\Gamma\,\theta_K\,\zeta_M\quad(C_6). \label{sm:eq:indsign}
\end{equation}
Reversing either convention (passive rotations or the opposite orientation of $C$) reverses the sign of the exponent and gives the form of Eq.~(33) of Ref.~\cite{Fang2012}. Equation~\eqref{sm:eq:indsign} agrees with the Fukui--Hatsugai--Suzuki Chern numbers of all 24 isolated bands of four test Hamiltonians.
\item \emph{$K$ and $M$: fingerprints of the CLS.} At $K$ and $M$, the Bloch form of the CLS does not vanish (Sec.~S6B), so the twin is the Bloch sum of its CLS. Rotations about the CLS center permute the translated copies without phases at a rotation-invariant momentum, so the eigenvalue is that of the CLS itself. The odd CLS alternates on all twelve ring bonds and is invariant under $C_6$, which shifts the ring by two sites, whereas the even CLS has the pattern $(+,-,-,+,\dots)$ and changes sign. The CLS eigenvalue is therefore $-\eta$, so $\theta_K=(-\eta)^2=1$ and $\zeta_M=(-\eta)^3=-\eta$. These values do not depend on $t'$, because the CLS does not, and cannot change under the perturbation as long as the twin stays nondegenerate at $K$ and $M$.
\item \emph{$\Gamma$: the doublet.} By Sec.~S2A, the two members of the doublet $|\chi=\pm1,\eta\rangle$ have $\xi_\Gamma=\eta\,\omega^\chi$, and by Sec.~S4 the chiral term shifts the member $\chi$ by $\chi m_\eta$. In the convention of Sec.~S0, this is angular momentum $\ell=\chi$ (mod 6) for the odd twin and $\ell=-2\chi$ (mod 6) for the even twin.
\item Steps 2--4 give $e^{-i\pi C_\eta/3}=(\eta\,\omega^{\chi_{\rm sel}})(1)(-\eta)=-e^{2\pi i\chi_{\rm sel}/3}=e^{-i\pi\chi_{\rm sel}/3}$. The dimer parity cancels, $C_\eta\equiv\chi_{\rm sel}$ (mod 6), and by step 1, $C_\eta=\chi_{\rm sel}$.
\item \emph{Selection.} For weak $m_\eta$ the gapped pair is $E_\eta+\tfrac12c_\eta k^2\pm(\tfrac14c_\eta^2k^4+m_\eta^2)^{1/2}$ near $\Gamma$. The branch that approaches $E_\eta+c_\eta k^2$ at large $k$ is the partner, which swallows the member on its side. The branch that approaches $E_\eta$ is the twin, which keeps the member at $E_\eta-|m_\eta|\,\mathrm{sgn}(c_\eta)$, the one on its own side of the partner. Hence $\chi_{\rm sel}=-\mathrm{sgn}(c_\eta)\,s_\eta$, and Eq.~\eqref{sm:eq:sigmabk} gives $C_\eta=\eta\kappa_\eta s_\eta$. $\square$
\end{enumerate}

\emph{Remarks.}
\begin{enumerate}
\item \emph{Inversion eigenvalues.} The $C_2=C_6^3$ eigenvalues of the twin give only $(-1)^C=-1$, i.e., $C$ modulo two, consistent with the inversion argument of Ref.~\cite{Ma2020}. The sign resides in the $C_3$ eigenvalue $\xi_\Gamma^2=\omega^{-\chi}$.
\item \emph{What the graph supplies.} The indicator converts the rotation eigenvalues of an isolated band into $C$ modulo six, but it does not say which eigenvalues the band acquires, and it cannot be applied to the twin before the gap opens, because two states with different $C_6$ eigenvalues are degenerate at $\Gamma$. The theorem supplies these eigenvalues from graph structure: at $K$ and $M$ from the CLS (step 3), at $\Gamma$ from the doublet and from the member selected by the mass and the side of the partner (steps 4 and 6). Step 1 lifts the result from modulo six to the integer.
\item \emph{$C_6$-breaking deformations.} Since $C$ cannot change without closing the isolating gap, Eq.~\eqref{sm:eq:thm} also fixes the Chern number of any $C_6$-breaking deformation (e.g., unequal $t'$ on the three dimers) that is connected to the symmetric lattice without closing the twin's gap.
\end{enumerate}

\subsection{Inside the triple point: the $A_1$ level hands over the Chern number}\label{sm:sec:handover}
For $|2t'-dt|\lesssim|m_{\rm even}|$ three levels are close at $\Gamma$: the even doublet, split into two members, and the $A_1$ level $|A_1,-\rangle$. The $A_1$ level is real, so the $C_2\mathcal T$-odd chiral terms have no expectation value in it, and its $C_6$ eigenvalue $-1$ differs from those of the doublet, so no $C_6$-symmetric term couples it to the doublet, and it moves only through the real parts of the chiral terms. When it lies between the two split members, the even twin, band 3, holds the $A_1$ level at $\Gamma$, with $\xi_\Gamma=-1$, and steps 3 and 5 give $e^{-i\pi C/3}=(-1)(1)(-1)=1$, so $C\equiv0$ (mod 6). For weak chiral terms the Berry flux of band 3 is confined to a neighborhood of $\Gamma$ of size $\sim|m|^{1/2}$ and $|C|\le1$, hence $C=0$. The rule thus extends to the triple point in the form: \emph{the twin carries the chirality of the state it holds at $\Gamma$}, which inside the window is the nonchiral $A_1$ level.

\begin{figure}[h]
\includegraphics[width=\textwidth]{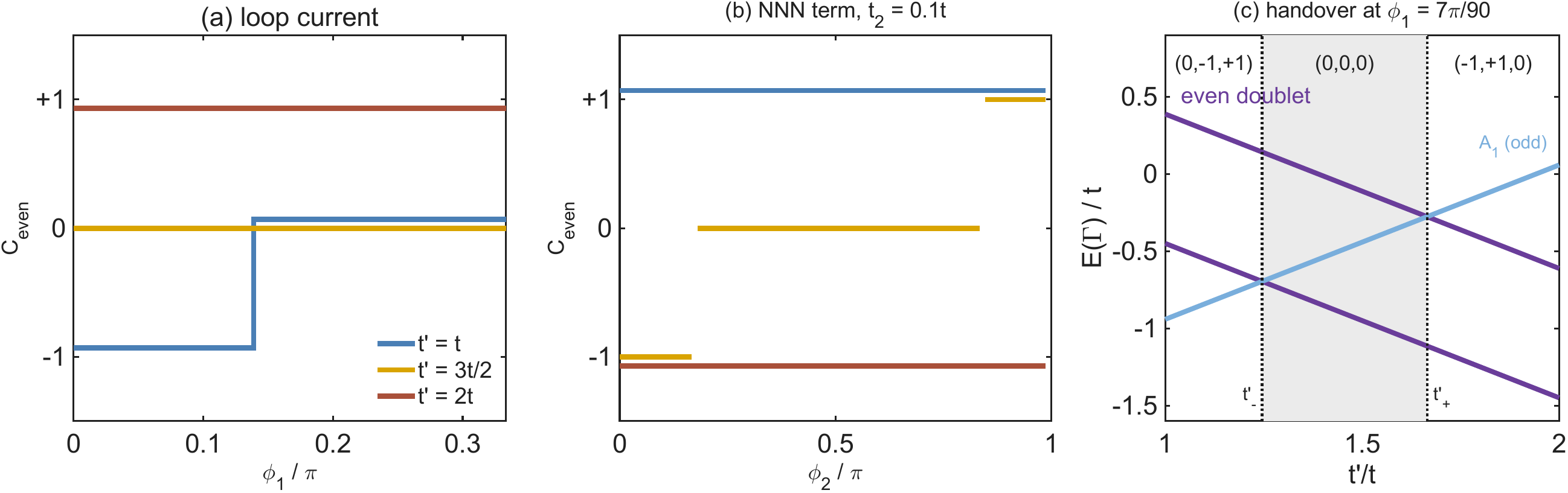}
\caption{\label{sm:fig:Shandover}The even twin changes partner. (a)~Chern number of the even twin versus the loop current $\phi_1$ in the star-lattice type regime ($t'=t$, blue), at the triple point ($t'=3t/2$, yellow), and in the double-kagome type regime ($t'=2t$, brown). The twin reverses across the triple point and is trivial at it. At $t'=t$ it becomes trivial beyond $\phi_1^c=0.1374\pi$, where the $A_1$ level enters the split doublet. (b)~The same for the NNN term ($t_2=0.1t$) versus $\phi_2$. At $t'=3t/2$ the real part of the NNN term shifts the window off the triple point for small $|\sin\phi_2|$. The odd twin keeps $C=+1$ in all cases. (c)~The six $\Gamma$ levels near the even twin at $\phi_1=7\pi/90$: the $A_1$ level of odd parity (light blue) rises through the split even doublet (purple) at $t'_\pm=\sqrt3\,t\cos(\phi_1\mp\pi/6)$ (dotted). The labels give $(C^{(4)},C^{(3)},C^{(2)})$ in the three intervals, and the shaded interval is the trivial window.}
\end{figure}

\emph{Exact window for the loop current.} Because the $\Gamma$ levels are exact diagonal elements, the window follows in a closed form. With the loop current alone, the even doublet lies at $t\cos\phi_1\pm\sqrt3\,t\sin\phi_1-t'$, and the $A_1$ level at $-2t\cos\phi_1+t'$ (Sec.~S4). The $A_1$ level crosses the two members at
\begin{equation}
t'_\pm=\sqrt3\,t\cos(\phi_1\mp\pi/6), \label{sm:eq:window}
\end{equation}
so the window has width $t'_+-t'_-=\sqrt3\,t\sin\phi_1=|m_{\rm even}|$ and center $\tfrac d2t\cos\phi_1$, which returns to the triple point $3t/2$ as $\phi_1\to0$. At $\phi_1=7\pi/90$, Eq.~\eqref{sm:eq:window} gives $t'_-=1.246t$ and $t'_+=1.665t$, and a fine scan finds the gap closings of bands 2--3 and 3--4 at $1.246t$ and $1.665t$. The three bands around the even twin hand the Chern number over in two steps [Fig.~\ref{sm:fig:Shandover}(c)]: $(C^{(4)},C^{(3)},C^{(2)})=(0,-1,+1)$ below $t'_-$, $(0,0,0)$ between, and $(-1,+1,0)$ above $t'_+$. The rising $A_1$ level first meets the lower member of the doublet and takes over the Chern number of band 2, then meets the upper member and passes it on to band 4. With a chiral term present the pristine pseudospin-1 crossing thus splits into two separate twofold closings, and their sum keeps the total Chern number of the three bands fixed. For the NNN term the real part $2t_2\cos\phi_2$ shifts the doublet, so the window is centered at $dt/2$ only for $\phi_2=\pi/2$. At $t'=3t/2$ and $t_2=0.1t$ the even twin is trivial only for $\pi/6<\phi_2<5\pi/6$. For smaller $|\sin\phi_2|$ the window is narrower than this shift, the twin holds a chiral member at $\Gamma$ ($C_6$ eigenvalue $\omega^{-1}$), and it carries $C=\mp1$ [Fig.~\ref{sm:fig:Shandover}(b)].

Table~\ref{sm:tab:window} compares the window widths obtained from grid-gap closings with the exact $\Gamma$ crossings and $|m_{\rm even}|$. In the window the even twin is trivial and its flatness peaks ($f_3=11.3$ at $t'=1.50t$ for the NNN term and $12.6$ at $1.46t$ for the loop current), which is not an instance of the theorem of Ref.~\cite{Chen2014}, since that theorem concerns exactly flat bands. Near the triple point, local refinement of the $\mathbf k$ grid is essential: the Berry curvature of the even twin is concentrated within $k\sim\sqrt{|m|}$ of $\Gamma$, and a uniform $36^2$ grid returns a spurious $C^{(3)}=\pm1$ at $t'=3t/2$.
\begin{table}[h]
\caption{\label{sm:tab:window}The $C^{(3)}=0$ window. Edges (gap closings of band 3 with bands 2 and 4) on a $24^2$ grid in steps of $\Delta t'=0.005t$. Width of the window from these edges, from the exact $\Gamma$-point crossings, and $|m_{\rm even}|$ [$\sqrt3\,t\sin\phi_1$ for the loop current, $2\sqrt3\,t_2$ for the NNN term at $\phi_2=\pi/2$].}
\begin{ruledtabular}
\begin{tabular}{llccccc}
term & strength & $3|2$ & $3|4$ & width (grid) & width (exact $\Gamma$) & $|m_{\rm even}|/t$ \\ \hline
loop & $\phi_1=0.01\pi$ & 1.470 & 1.525 & 0.055 & 0.0544 & 0.0544 \\
loop & $\phi_1=0.02\pi$ & 1.445 & 1.550 & 0.105 & 0.1088 & 0.1088 \\
loop & $\phi_1=0.05\pi$ & 1.345 & 1.615 & 0.270 & 0.2710 & 0.2710 \\
loop & $\phi_1=0.10\pi$ & 1.160 & 1.695 & 0.535 & 0.5352 & 0.5352 \\
NNN & $t_2=0.025t$ & 1.455 & 1.545 & 0.090 & 0.0866 & 0.0866 \\
NNN & $t_2=0.05t$ & 1.415 & 1.585 & 0.170 & 0.1732 & 0.1732 \\
NNN & $t_2=0.1t$ & 1.325 & 1.675 & 0.350 & 0.3464 & 0.3464 \\
NNN & $t_2=0.2t$ & 1.155 & 1.845 & 0.690 & 0.6928 & 0.6928 \\
\end{tabular}
\end{ruledtabular}
\end{table}

\subsection{Tests}\label{sm:sec:tests}
\begin{figure}[h]
\includegraphics[width=0.9\textwidth]{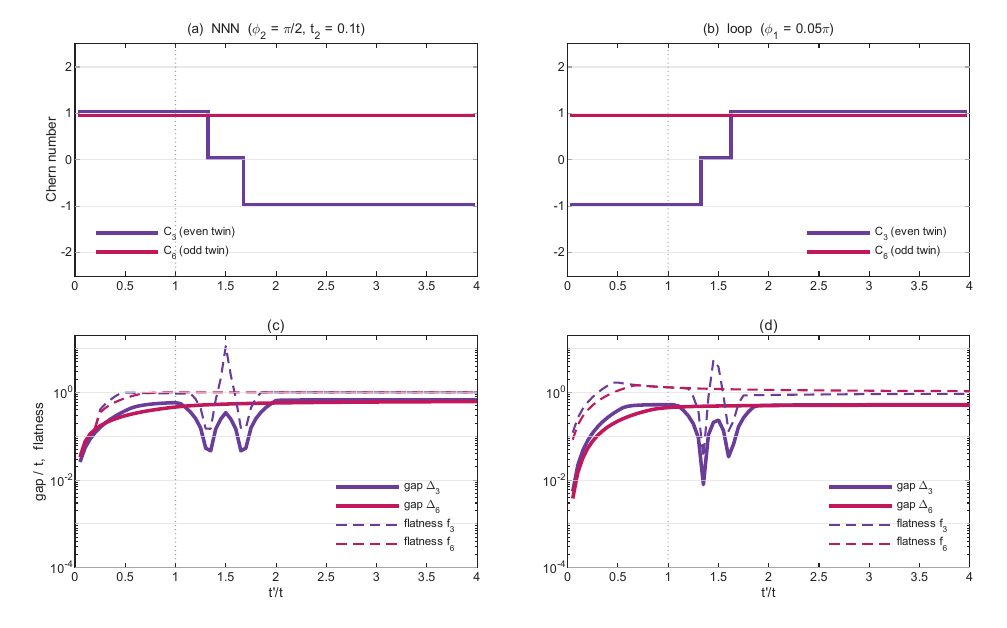}
\caption{\label{sm:fig:Stwin}Twin flat bands versus $t'/t$. Top: Chern numbers of the even (band 3) and odd (band 6) twins. Bottom: isolating gaps (solid) and flatness ratios from the indirect gap (dashed). Left: NNN term ($\phi_2=\pi/2$, $t_2=0.1t$). Right: loop current ($\phi_1=0.05\pi$). Purple: even twin. Raspberry: odd twin.}
\end{figure}

\emph{Single chiral terms versus $t'$.} Figure~\ref{sm:fig:Stwin} shows the Fukui--Hatsugai--Suzuki Chern numbers ($30^2$ grid), isolating gaps, and flatness ratios of the twins for $0.05\le t'/t\le4$. For the NNN term ($\phi_2=\pi/2$, $t_2=0.1t$) the band Chern numbers are $(+1,-1,+1,-2,0,+1)$ for $t'<1.32t$, $(+1,0,0,-2,0,+1)$ in the window, and $(+1,0,-1,-1,0,+1)$ for $t'>1.68t$. For the loop current ($\phi_1=0.05\pi$) they are $(-1,+1,-1,0,0,+1)$ for $t'<1.35t$, $(-1,0,0,0,0,+1)$ in the window, and $(-1,0,+1,-1,0,+1)$ for $t'>1.62t$. The odd twin has $C^{(6)}=+1$ throughout. At the parameters of Fig.~2 of the main text ($\phi_1=7\pi/90$ for the loop current, and $t_2=0.1t$, $\phi_2=\pi/2$ for the NNN term), a calculation on a $48^2$ grid with every plaquette within $|\mathbf k|<0.25$ of $\Gamma$ subdivided $40\times40$ gives $(-1,+1,-1,0,0,+1)$, $(-1,0,0,0,0,+1)$, and $(-1,0,+1,-1,0,+1)$ for the loop current at $t'=t$, $3t/2$, and $2t$, and $(+1,-1,+1,-2,0,+1)$, $(+1,0,0,-2,0,+1)$, and $(+1,0,-1,-1,0,+1)$ for the NNN term. These do not depend on the chosen strength: along $t'=t$ and $2t$ the same $(C_{\rm even},C_{\rm odd})$ hold from $|m_{\rm even}|\simeq0.02t$ up to the entry of the $A_1$ level into the even doublet, at $|m_{\rm even}|=0.725t$ for the loop current at $t'=t$ and at $1.00t$ for the NNN term at $t'=t$ and $2t$. For the loop current at $t'=2t$ the $A_1$ level stays outside over the whole scan ($|m_{\rm even}|\le1.07t$). The parameters of Fig.~2, $|m_{\rm even}|=0.42t$ and $0.35t$, lie well inside these ranges.

\begin{figure}[h]
\includegraphics[width=\textwidth]{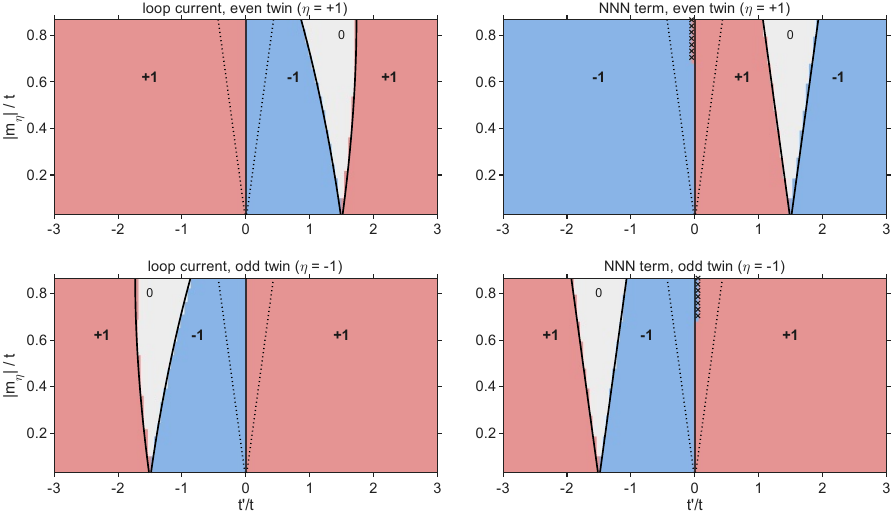}
\caption{\label{sm:fig:Sphase}The Chern rule over the $(t',|m_\eta|)$ plane. Colors: Fukui--Hatsugai--Suzuki Chern number of the twin (blue $-1$, gray $0$, red $+1$) for the loop current (left) and the NNN term (right), even twin (top) and odd twin (bottom). Solid lines: the boundaries predicted by the rule, i.e., the edges of the trivial window around the triple point $2\eta t'=dt$ [Eq.~\eqref{sm:eq:window} for the loop current] and $t'=0$, where $\mathrm{sgn}(\eta t')$ changes sign. Dotted lines: $|m|=2|t'|$, where the twins meet. Crosses: points where the rule and the computed Chern number disagree, all in the region where the twins are mixed. Chern numbers on a $36^2$ grid, with $|m_\eta|=\sqrt3\,t\sin\phi_1$ for the loop current and $2\sqrt3\,t_2$ at $\phi_2=\pi/2$ for the NNN term.}
\end{figure}

\emph{The full map.} Figure~\ref{sm:fig:Sphase} compares Eq.~\eqref{sm:eq:thm} with Fukui--Hatsugai--Suzuki Chern numbers over $t'/t\in[-3,3]$ (step $0.05$) and $|m_\eta|\le0.87t$, for both chiral terms and both twins. All points with $|m_\eta|<2|t'|$, so that the two twins stay apart, obey $C_\eta=\mathrm{sgn}(\eta t')\,\kappa_\eta s_\eta$, Eq.~\eqref{sm:eq:thm} with $\eta\to\mathrm{sgn}(\eta t')$, when the $A_1$ level lies outside the split doublet and $C=0$ when it lies between its members, as read from the exact $\Gamma$ levels (3294 of 3294 points per panel). The remaining points, at $|t'|\le0.05t$ and $|m_\eta|\gtrsim0.7t$, have the two twins mixed. For $t'<0$ the twins exchange roles, as the map $\Sigma$ requires.

\begin{figure}[h]
\includegraphics[width=0.95\textwidth]{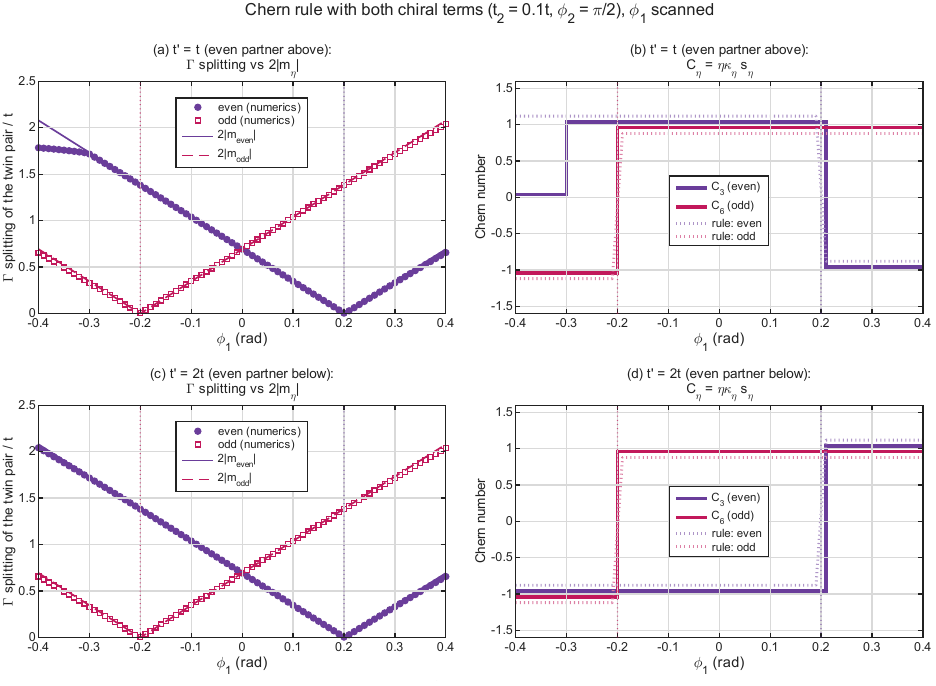}
\caption{\label{sm:fig:Srule}The Chern rule with both chiral terms present ($t_2=0.1t$, $\phi_2=\pi/2$, $\phi_1$ scanned).
(a),(c)~$\Gamma$ splitting of the even (purple) and odd (raspberry) twin pairs at $t'=t$ and $t'=2t$, compared with $2|m_\eta|$.
(b),(d)~Chern numbers $C^{(3)}$ (even) and $C^{(6)}$ (odd) compared with $C_\eta=\eta\kappa_\eta s_\eta$ (dotted). The value $C^{(3)}=0$ at $\phi_1\le-0.31$ in (b) is the case of Sec.~\ref{sm:sec:handover}, with the $A_1$ level between the split members. Between (b) and (d), the even twin crosses the triple point, so $\kappa_{\rm even}$ and hence the even-twin Chern numbers are reversed, while the odd twin is unchanged.}
\end{figure}

\emph{Both chiral terms.} The theorem covers the single-term results and the reversal of the even twin at the triple point. A direct test with both terms present is shown in Fig.~\ref{sm:fig:Srule} ($t_2=0.1t$, $\phi_2=\pi/2$, $\phi_1\in[-0.4,0.4]$ rad, at $t'=t$ and $2t$). The $\Gamma$ splittings of the twin pairs follow $2|m_\eta|$ (maximal deviation $0.037t$, except for the even twin at $t'=t$ and $\phi_1<-0.3$). Each twin closes its gap only on its own compensation line, $m_\eta=0$, i.e., $t\sin\phi_1=2\eta\,t_2\sin\phi_2$: $t\sin\phi_1=+0.2t$ for the even twin and $-0.2t$ for the odd twin, so the two Chern numbers can be set independently. With the position of the $A_1$ level read from the exact $\Gamma$ levels, Eq.~\eqref{sm:eq:thm} and the triple-point statement reproduce the Chern numbers at all gapped points ($4\times81$ points).

\subsection{Another root: the square-lattice analog $L(S(\text{square}))$}\label{sm:sec:square}
\begin{figure}[h]
\includegraphics[width=0.95\textwidth]{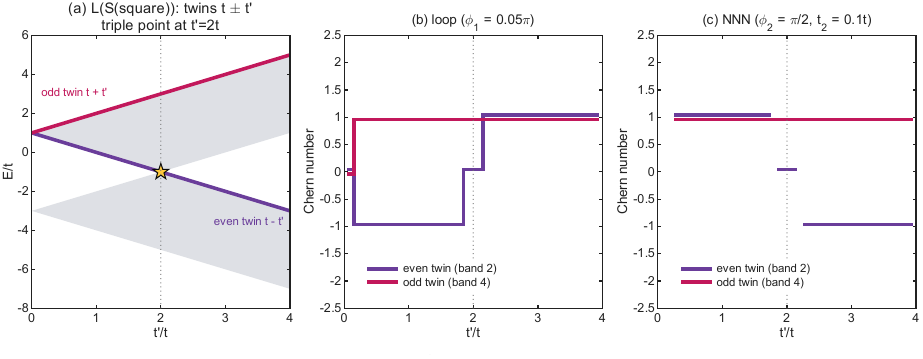}
\caption{\label{sm:fig:Ssquare}Square-lattice analog $L(S(\text{square}))$.
(a)~Band ranges and twins $t\pm t'$ versus $t'/t$, with the triple point at $t'=2t=dt/2$.
(b)~Chern numbers of the even (band 2) and odd (band 4) twins for the clique loop current, $\phi_1=0.05\pi$.
(c)~Same as (b) except this panel refers to the NNN term, $\phi_2=\pi/2$, $t_2=0.1t$. Gaps in the curves mark gap closings.}
\end{figure}

To test which statements are properties of the class, we repeat the analysis for $G$, the square lattice ($d=4$), with four sites per cell and four-site cliques $K_4$. The loop current $te^{i\phi_1\nu}$ acts on the four perimeter bonds of each clique, oriented counterclockwise, with real diagonals, and the NNN term $t_2e^{i\phi_2\nu}$ on every two-step path of one perimeter bond and one dimer bond (Fig.~\ref{sm:fig:Ssquare}).
\begin{enumerate}
\item \emph{Twins and triple point.} The twins lie at $t\pm t'$. The odd twin touches a dispersive band at $\Gamma$ and the even twin at $M$, as required for a bipartite root. At $M$ the even twin is degenerate with two dispersive levels at $t'=2t=dt/2$, e.g., $E(M)=\{-1,-1,-1,3\}t$.
\item \emph{Parity identity.} Projected onto the twin doublet, the NNN block equals $-2\eta$ times the loop-current block for $t'/t=0.5$, $1$, and $3$ (residual at machine precision).
\item \emph{Chern pattern.} At $t'=t$ the loop current gives $C_{\rm even}=-1$ and $C_{\rm odd}=+1$, whereas the NNN term gives $C_{\rm even}=C_{\rm odd}=+1$, exactly as on the SL. Across the triple point the even twin reverses, through a $C=0$ window centered at $t'=2t$, whereas the odd twin is unaffected.
\item \emph{Flatness pinning.} For weak chiral terms at $t'=t$, $f=1.00$ (NNN, $t_2\le0.02t$) and $1.00\pm0.03$ (loop, $\phi_1\le0.01\pi$). At the triple point, where the even twin is trivial, $f$ reaches $12$ for the NNN term.
\end{enumerate}

\emph{Proof for $C_4$.} The theorem holds with $\eta\kappa_\eta=+1$ for the odd twin and $\mathrm{sgn}(t'-2t)$ for the even twin. We use active rotations about the plaquette center of $G$. The sites of a clique are labeled $j=0,1,2,3$ (at $0^\circ$, $90^\circ$, $180^\circ$, and $270^\circ$ about its vertex), and the clique orbitals are $|\ell\rangle=\frac12\sum_j i^{\ell j}|j\rangle$. Sites $j$ and $j+2$ of neighboring cliques are dimer partners, and the chirality $\chi=\pm1$ is the circulation $\ell=\chi$ around the clique perimeter. Steps 1 and 6 are unchanged, with the touching at $\Gamma$ for the odd twin and at $M$ for the even twin. The other steps become:
\begin{enumerate}
\item[(2)] \emph{Indicator.} Equation~\eqref{sm:eq:indsign} with $n=4$ gives $i^{-C}=\xi_\Gamma\,\xi_M\,\zeta_X$, with $\xi$ the $C_4$ eigenvalues at $\Gamma$ and $M$ and $\zeta_X$ the $C_2$ eigenvalue at $X$. It agrees with the Fukui--Hatsugai--Suzuki Chern numbers of all eight gapped twins.
\item[(3)] \emph{CLS momenta.} Away from its touching point, at $X$ and at $M$ (odd) or $\Gamma$ (even), the twin is the Bloch sum of its CLS on a plaquette ring of eight sites. The odd CLS alternates around the ring, and the even CLS has the pattern $(+,+,-,-,\dots)$. $C_4$ shifts the ring by two sites, so the CLS eigenvalue is $-\eta$, as on the SL: $\xi_M=+1$ for the odd twin, $\xi_\Gamma=-1$ for the even twin, and $\zeta_X=1$ for both.
\item[(4)] \emph{Doublet at the touching point.} At $\Gamma$ (odd) and $M$ (even), dimer parity together with the Bloch phase $\pm1$ requires $x_{j+2}=-x_j$ on the clique, so only $|\ell=\pm1\rangle$ survive, both at $t-\eta t'$. A $90^\circ$ rotation about the vertex gives $i^{-\chi}$, and the rotation about the plaquette center differs from it by a translation, which contributes $1$ at $\Gamma$ and $-1$ at $M$. Hence $\xi_\Gamma=i^{-\chi}$ for the odd twin and $\xi_M=-i^{-\chi}$ for the even twin. The chiral terms are diagonal in $\chi$ and shift the member $\chi$ by $\chi m_\eta$.
\item[(5)] \emph{Result.} For the odd twin $\xi_\Gamma\xi_M\zeta_X=i^{-\chi_{\rm sel}}$, and for the even twin $(-1)(-i^{-\chi_{\rm sel}})(1)=i^{-\chi_{\rm sel}}$. Hence $C_\eta\equiv\chi_{\rm sel}$ (mod 4), and by step 1, $C_\eta=\chi_{\rm sel}=\eta\kappa_\eta s_\eta$. As on the SL, $\eta$ cancels between the touching point and the CLS momenta. At $t'=2t$ the $A_1$ orbital $|\ell=0\rangle$ joins the even doublet at $M$, with eigenvalue $-1$. When it lies between the split members the indicator gives $(-1)(-1)(1)=1$, and $C=0$, as observed.
\end{enumerate}

\emph{Other roots.} The ingredients of the rule separate. Equation~\eqref{sm:eq:closed} and Sec.~\ref{sm:sec:side} hold for any connected $d$-regular root, so the bonding sign $\eta$, the dominance sign $\kappa_\eta$, and the side of the partner, Eq.~\eqref{sm:eq:sigmabk}, are fixed by the graph in general. The proof of Sec.~\ref{sm:sec:proof} needs, in addition, a rotation that maps each clique onto itself up to a translation and gives the two members of the doublet different eigenvalues. This rotation fixes the double winding of the touching in step 1 and the eigenvalues in steps 2--5. We have carried out the proof for $C_6$ (the SL) and $C_4$ (above). Where a twin consists of several bands, $N_e-N_v>1$, as for the triangular root (Table~\ref{sm:tab:graph}), the proof does not apply as it stands.

\section{Flatness}

\subsection{Flatness pinning of a gapped singular flat band}\label{sm:sec:pinning}
\emph{Physical picture.} Away from its touching point, the twin is spanned by its CLS, which are real. A time-reversal-odd term is purely imaginary in this real basis, so it cannot shift a real state to first order. It acts only where the twin meets its partner, at $\Gamma$, and there it splits the pair symmetrically. The gapped twin therefore keeps a bandwidth equal to its own gap.

\emph{Statement.} Let $H_0$ be $C_2\mathcal T$ symmetric, with a singular flat band at $E_0$ that touches a dispersive band only at $\Gamma$. Let the perturbation be $V=V_R+V_I$, where $V_R$ is $C_2\mathcal T$ even and preserves the CLS exactly, and $V_I$ is $C_2\mathcal T$ odd. Then, to first order, the flat band stays flat away from $\Gamma$, and the touching splits symmetrically about it. Consequently the bandwidth $W$ and the isolating gap $\Delta_{\rm iso}$ both equal $|m|$, and
\begin{equation}
f=\Delta_{\rm iso}/W=1+O(|m|/t).
\end{equation}

\emph{Proof.}
\begin{enumerate}
\item In a $C_2\mathcal T$-real basis, $H_0(\mathbf k)$ is real and the flat-band eigenvector $u_{\rm FB}(\mathbf k)$ can be chosen real. $V_I(\mathbf k)$ is purely imaginary and Hermitian, so $\langle u_{\rm FB}|V_I|u_{\rm FB}\rangle=0$ for every $\mathbf k$.
\item $V_R$ maps each CLS onto itself with a constant shift $c$. For the NNN term, $c=-2t_2\cos\phi_2$ ($+2t_2\cos\phi_2$) for the odd (even) twin. Away from $\Gamma$ the first-order energy is therefore the $\mathbf k$-independent constant $E_0'=E_0+c$.
\item At $\Gamma$ the flat band is degenerate with the partner. On this two-dimensional space $V_R=c\,\openone$, because $V_R$ alone preserves the touching, whereas $V_I=m\,\tau_z$ is traceless (Sec.~S4). The touching splits into $E_0'\pm m$.
\item The flat band then spans $[E_0',E_0'+|m|]$ (or $[E_0'-|m|,E_0']$) to first order, and the partner starts at the other split level. Hence $W=\Delta_{\rm iso}=|m|$. $\square$
\end{enumerate}
The statement uses only a real parent, a single twofold touching with one partner, and a first-order term odd under $C_2\mathcal T$. It therefore applies to any $C_2\mathcal T$-symmetric singular flat band with a twofold touching, including the line- and split-graph flat bands of Ref.~\cite{Ma2020} with spin-orbit coupling in each spin sector. It does not apply where three levels meet, at the triple point, where the trivial even twin reaches $f\simeq11$--$13$ (Sec.~S5B).

\emph{Numerical checks.} $\max_{\mathbf k}|\langle u_{\rm FB}|V_I|u_{\rm FB}\rangle|<4\times10^{-13}$ for both twins and both chiral terms. $W$ and $\Delta_{\rm iso}$ of both twins equal $|m_\eta|$ to first order at $t'=t$ and $2t$. For example, for the NNN term with $t_2=0.08t$ at $t'=t$, $W=\Delta_{\rm iso}=0.2771t=|m_{\rm odd}|$ for the odd twin and $W=0.2835t$, $\Delta_{\rm iso}=0.2708t$ for the even twin. The flatness ratio of the odd twin in the NNN model is $f=1.0000$ for $t_2\le0.05t$ and $0.998$ at $t_2=0.1t$ ($\phi_2=0.3\pi$), and that of the even twin in the loop model is $1.028$, $1.056$, $1.115$, and $1.31$ at $\phi_1/\pi=0.005$, $0.01$, $0.02$, and $0.05$, a linear correction. Outside the class $L(S(G))$, the kagome flat band with a Haldane-type NNN term gives $f=1.0000$ for strengths $0.005$--$0.04t$.

\emph{Consequence.} A weak chiral term cannot make a gapped twin flatter than $f\simeq1$, and any improvement is a higher-order effect. Over the full range of the loop model we find $f\le1.47$ for the Chern bands at $t'=t$ and $f\le1.82$ at $t'/t=0.8$ (scan over $t'/t\in[0.2,1.4]$, $\phi_1$ in $1^\circ$ steps). Large $f$ occurs only for trivial bands: in the triple-point window (Sec.~S5B) and near the exact flat bands of Sec.~S6B.

\subsection{Ring compact localized states and the exact flat-band ladder}\label{sm:sec:ladder}
\emph{Physical picture.} A loop current does not destroy every compact state. A state on one dodecagon ring stays trapped if the waves leaking from the two ring sites of each triangle into its third vertex cancel, and if the phase picked up around the ring closes on itself. The first condition fixes the phase jump on each triangle bond, and the second is an Aharonov--Bohm condition on the triangle flux~\cite{Vidal1998,Pal2025}. The two are commensurate only at discrete fluxes, where an exactly flat, and therefore trivial, band appears.

\emph{Derivation.} Label the ring sites counterclockwise as $a_1,b_1,\dots,a_6,b_6$, with $(a_j,b_j)$ in triangle $j$ and $(b_j,a_{j+1})$ joined by a dimer. Traversing the ring counterclockwise runs clockwise around each triangle, so $H_{a_jb_j}=-te^{-i\phi_1}$. Let $v_j$ be the vertex of triangle $j$ off the ring. No leakage onto $v_j$ requires
\begin{equation}
\psi_{b_j}=-e^{-2i\phi_1}\,\psi_{a_j}, \label{sm:eq:leak}
\end{equation}
a phase jump $\delta_{\rm in}=\pi-2\phi_1$ with equal moduli. By $C_6$ the jump across the dimers, $\psi_{a_{j+1}}=e^{i\delta_{\rm out}}\psi_{b_j}$, is uniform. The eigenvalue equations at $a_j$ and $b_j$ read $E=te^{-3i\phi_1}-t'e^{-i\delta_{\rm out}}$ and its complex conjugate, so $E$ is real only if $t'\sin\delta_{\rm out}=t\sin3\phi_1$, and single-valuedness requires $6(\delta_{\rm in}+\delta_{\rm out})\in2\pi\mathbb Z$. For $t'=t$ there are two branches. Branch (i), $\delta_{\rm out}=3\phi_1$, gives $E=0$ and requires $\phi_1\in(\pi/3)\mathbb Z$: it is the even CLS of the pristine lattice. Branch (ii), $\delta_{\rm out}=\pi-3\phi_1$, gives
\begin{equation}
E=2t\cos\Phi,\qquad \Phi\equiv3\phi_1=\frac{n\pi}{5},\ n\in\mathbb Z , \label{sm:eq:quant}
\end{equation}
with $\Phi$ the triangle flux. At $\Phi=0$ this is the odd CLS. Within one period the ladder has exactly flat bands at $\phi_1=\pi/15$, $2\pi/15$, $\pi/5$, and $4\pi/15$, with the golden-ratio energies $E/t=(1+\sqrt5)/2$, $(\sqrt5-1)/2$, $-(\sqrt5-1)/2$, and $-(1+\sqrt5)/2$ (bands 5, 4, 3, and 2). A scan $\phi_1=k\pi/60$ on a finite cluster finds ring CLS only at these phases. There the band width is below $10^{-8}t$, and the amplitudes have equal moduli and the predicted phase jumps.

\emph{Bloch form and the Chen theorem.} The Bloch sum $q(\mathbf k)$ of a CLS is a Laurent polynomial in $e^{i\mathbf k\cdot\mathbf a_{1,2}}$ with $H(\mathbf k)q=Eq$ for all $\mathbf k$. Chen \emph{et al.}~\cite{Chen2014} show that an isolated exactly flat band of a strictly local model has such a section that is nowhere vanishing after removal of common factors, which forces $C=0$. For the pristine twins $q$ vanishes at $\Gamma$, the defining property of a singular flat band: the CLS become linearly dependent there, and the partner supplies the missing state~\cite{Rhim2019,Bergman2008}. The cancellation has a simple picture. Each CLS is an oriented loop whose amplitudes reverse with the direction of traversal, and every edge of $G$ borders two faces traversed in opposite directions, so the equal-weight sum of all CLSs, which is $q$ at $\Gamma$, cancels for any $G$. Away from $\Gamma$, in particular at $K$ and $M$, $q\neq0$, which step 3 of Sec.~S5A uses. For the ladder bands $\min_{\mathbf k}\|q\|^2=7.15$ ($\phi_1=\pi/15$) and $8.29$ ($2\pi/15$), so they are isolated and trivial, which the Fukui--Hatsugai--Suzuki calculation confirms. Their Berry curvature is nonzero pointwise ($\max|\Omega|=0.92$ and $2.57$). Conversely, a twin with $C\neq0$ cannot be exactly flat anywhere in the region where the rule gives $C\neq0$.

\begin{figure}[h]
\includegraphics[width=\textwidth]{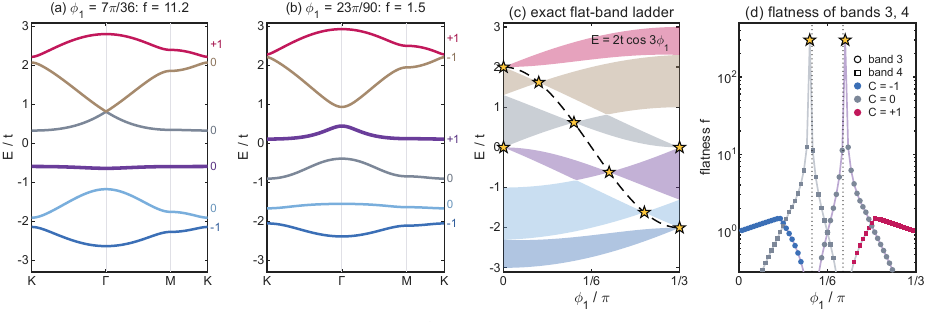}
\caption{\label{sm:fig:Sflat}Flatness in the loop model at $t'=t$.
(a),(b)~Bands at $\phi_1=7\pi/36$ and $23\pi/90$, with band Chern numbers labeled at the right end of each band. The highlighted band near $E=0$ is trivial with $f=11.2$ in (a) (band 3) and has $C=+1$ with $f=1.5$ in (b) (band 4).
(c)~Band ranges versus $\phi_1$ and the exact flat-band ladder $E=2t\cos3\phi_1$ at $\Phi=3\phi_1=n\pi/5$ (stars), Eq.~\eqref{sm:eq:quant}.
(d)~Flatness ratios of bands 3 (circles) and 4 (squares), colored by their own Chern numbers (blue $C=-1$, gray $C=0$, raspberry $C=+1$). Dotted lines mark the Chern transfers at $\phi_1^c$ and $\pi/3-\phi_1^c$, and the exact flat bands (stars) lie in the trivial intervals of their own bands.}
\end{figure}

\emph{Where the ladder sits.} The spectrum of the loop model is $2\pi/3$ periodic in $\phi_1$, and the mirror $\phi_1\to\pi/3-\phi_1$ maps $E\to-E$ and $C^{(n)}\to-C^{(7-n)}$. Band 3 is trivial beyond $\phi_1^c=0.1374\pi$, where the $A_1$ level enters the split even doublet, and each exact flat band lies in the trivial interval of its own band. The trivial bands near the ladder reach $f=11.2$ ($\phi_1=7\pi/36$), whereas the flattest Chern bands, at $\phi_1=7\pi/90$ (band 3, $C=-1$) and its mirror $23\pi/90$ (band 4, $C=+1$), have $f=1.47$, as the pinning predicts (Fig.~\ref{sm:fig:Sflat}).

\section{Quantum geometry of a gapped twin}\label{sm:sec:geometry}
\emph{Physical picture.} The Berry flux of a gapped twin is not spread over the BZ. It sits on a ring around $\Gamma$, where the mass $m_\eta$ and the quadratic splitting $\tfrac12|c_\eta|k^2$ are comparable. The ring shrinks as the gap closes, while the quantum metric keeps a $1/k^2$ tail from the singular touching, so the integrated metric grows logarithmically.

\emph{Effective model.} Project onto the doublet $|{\pm1},\eta\rangle$ at $\Gamma$. $C_3$ forbids diagonal $\chi$-odd $k^2$ terms and allows the term $k^2e^{\pm2i\theta}$ that changes $\chi$ by two units. With one eigenvalue $\mathbf k$-independent,
\begin{equation}
H_{\rm eff}=E_\eta+\tfrac12c_\eta k^2+\tfrac12c_\eta k^2\,(\cos2\theta\,\tau_x+\sin2\theta\,\tau_y)+m_\eta\,\tau_z, \label{sm:eq:Heff}
\end{equation}
with $\tau_z=\chi$. Write $\mathbf d=(\rho\cos2\theta,\rho\sin2\theta,m)$, $\rho=|c|k^2/2$, and $\tilde c=|c|/2$. For the twin,
\begin{equation}
|\Omega(\mathbf k)|=\frac{2|m|\,\tilde c^{\,2}k^2}{(m^2+\tilde c^{\,2}k^4)^{3/2}}, \label{sm:eq:Sring}
\end{equation}
which vanishes as $k^2$ at $\Gamma$ and peaks on a ring of radius $k^*=2^{1/4}(|m|/|c|)^{1/2}$. Its integral is $1$, so the ring carries the full unit of flux. The metric is
\begin{equation}
\tr g=\tilde c^{\,2}k^2\Big[\frac{m^2}{(m^2+\rho^2)^2}+\frac{1}{m^2+\rho^2}\Big]\ \xrightarrow{\ k\gg k^*\ }\ \frac1{k^2},
\end{equation}
with unit coefficient for both twins. Its integral up to $u_c=\tilde c\Lambda^2$, with $\Lambda\sim1/a$ the scale where the two-band description fails, is $\mathcal G=\int\tr g\,d^2k/2\pi=\tfrac14+\tfrac14\ln[(m^2+u_c^2)/m^2]\simeq\tfrac12\ln(1/|m|)+O(1)$, the logarithmic enhancement known for singular flat bands~\cite{Jiang2025}. Since $\int|\Omega|=1$, the trace-condition violation is $T=\mathcal G-1$. The slope $\tfrac12$ is reached only for $k^*\ll\Lambda$. Because $|c_{\rm odd}|=|c_{\rm even}|/5$ at $t'=t$, the ring of the odd twin is $\sqrt5$ times larger at equal $|m|$, so the odd twin reaches the asymptotic regime later and stays closer to the bound $\mathcal G\ge|C|$.

\begin{figure}[h]
\includegraphics[width=\textwidth]{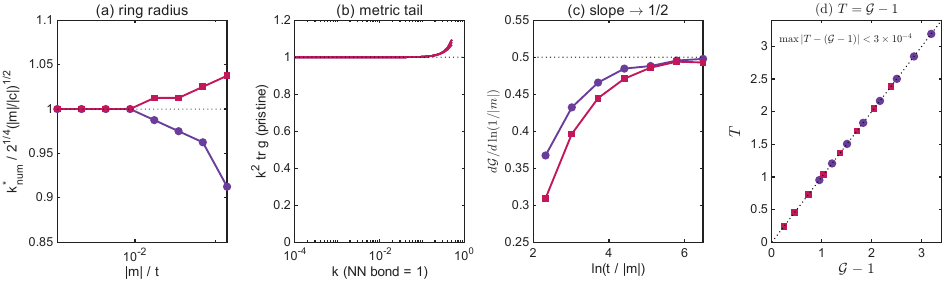}
\caption{\label{sm:fig:Stwoband}The Berry ring of a gapped twin: two-band theory versus the lattice (NNN term, $\phi_2=\pi/2$, $t'=t$, with the even twin as purple circles and the odd twin as raspberry squares).
(a)~Peak radius of $|\Omega|$ divided by $2^{1/4}(|m|/|c|)^{1/2}$.
(b)~$k^2\tr g$ of the pristine twins versus $k$ along four directions, which tends to 1 for both twins.
(c)~Local slope $d\mathcal G/d\ln(1/|m|)$, approaching $\tfrac12$, later for the odd twin.
(d)~$T$ versus $\mathcal G-1$. The dotted line is $T=\mathcal G-1$.}
\end{figure}

\emph{Lattice check} (NNN term, $\phi_2=\pi/2$, $t'=t$, $|m|=2\sqrt3\,t_2$ from $0.14t$ to $0.0011t$, physical orbital positions, and logarithmic radial quadrature near $\Gamma$). The peak radius of $|\Omega|$ agrees with $2^{1/4}(|m|/|c|)^{1/2}$, with $|c_{\rm even}|=3.482$ and $|c_{\rm odd}|=0.696$ (Sec.~S3), within $1.3\%$ for $|m|\le0.017t$. For the pristine twins $k^2\tr g=1.0000$ at $k=10^{-2}$ to $10^{-4}$. The local slope $d\mathcal G/d\ln(1/|m|)$ rises to $0.498$ (even) and $0.493$ (odd), and $T=\mathcal G-1$ within $3\times10^{-4}$ (Fig.~\ref{sm:fig:Stwoband}).

\section{Edge channels switched by a real coupling}\label{sm:sec:edge}
Edges and Hall transport see the Chern sum of all bands below a gap, not the Chern number of a single band.
Section~\ref{sm:sec:handover} shows that the even twin reverses in two steps, and each step closes only one of the two gaps around it.
The rising $A_1$ level first exchanges one unit of Chern number with the band below the twin and then one unit with the band above it.
Each gap Chern sum therefore changes by one, so a chiral edge channel switches on in one gap and off in the other, and no channel reverses its direction.
The reversal of the twin itself is seen only by band-resolved probes, such as the frequency-integrated circular dichroism of the twin alone or a local Chern marker restricted to it.

We take the loop current at $\phi_1=\pi/12$, where Eq.~\eqref{sm:eq:window} gives the trivial window $1.22t<t'<1.67t$, and the three values $t'/t=1$, $1.45$, and $2$.
All six bands are isolated by global gaps of at least $0.08t$.
The bands around the even twin carry $(C^{(4)},C^{(3)},C^{(2)})=(0,-1,+1)$, $(0,0,0)$, and $(-1,+1,0)$, and the odd twin keeps $C^{(6)}=+1$.
The Chern sums of the gaps below and above the even twin are thus $(0,-1)$, $(-1,-1)$, and $(-1,0)$.

\begin{figure}[h]
\includegraphics[width=\textwidth]{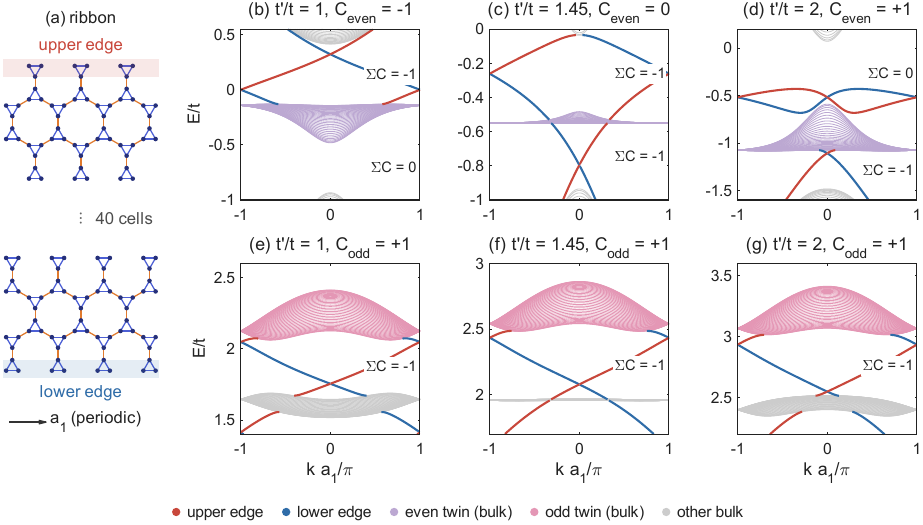}
\caption{\label{sm:fig:Sedge_ribbon}Edge channels on a ribbon (loop current $\phi_1=\pi/12$, 40 cells, periodic along $\mathbf a_1$). (a)~The ribbon, with its upper (red) and lower (blue) edges. (b)--(g)~Bands colored by where the states live: upper edge (red), lower edge (blue), bulk even twin (purple), bulk odd twin (raspberry), and other bulk states (gray), with the Chern sum $\Sigma C$ of each gap. (b)--(d)~Around the even twin at $t'/t=1$, $1.45$, and $2$. As $t'$ crosses the trivial window, the channel in the gap above the twin switches off and the one in the gap below switches on. In (d) the gap above the twin holds edge states that return to the same side of the gap and carry no net current. (e)--(g)~Around the odd twin, whose channel does not change.}
\end{figure}

\emph{Ribbon.} On a ribbon periodic along $\mathbf a_1$ and 40 cells wide, we count at the center of each gap the crossings of states with more than half of their weight in the four outermost cells, signed by their group velocity (Fig.~\ref{sm:fig:Sedge_ribbon}).
The signed count on the lower edge equals the gap Chern sum in all nine cases, the gap below the odd twin included.
In the gap above the even twin at $t'=2t$, the ribbon also holds edge states that leave and return to the same side of the gap, so their crossings cancel and they carry no net current.

\begin{figure}[h]
\includegraphics[width=\textwidth]{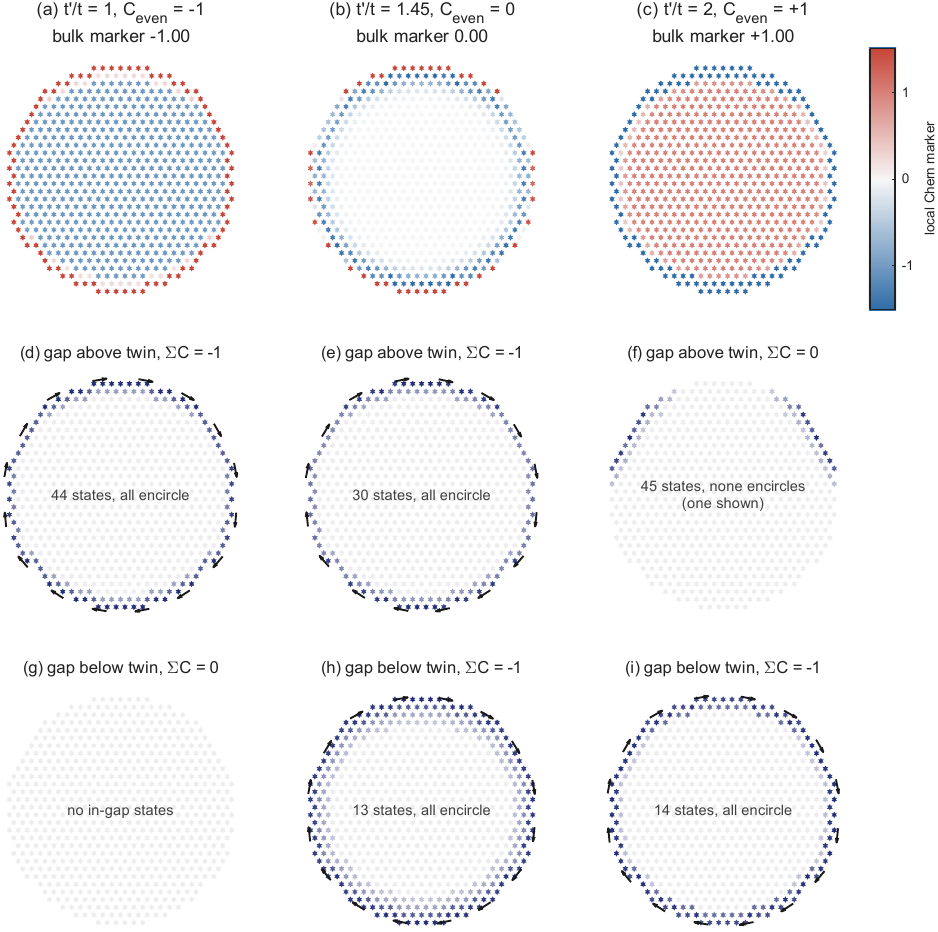}
\caption{\label{sm:fig:Sedge_flake}The twin reverses, the channels switch. (a)--(c)~Local Chern marker of the even twin on a flake of 613 cells at $t'/t=1$, $1.45$, and $2$, which is $-1$, $0$, and $+1$ in the bulk. (d)--(i)~In-gap states in the gaps above (d--f) and below (g--i) the twin, with the Chern sum $\Sigma C$ of each gap. Shading gives the weight of the in-gap states and arrows their current, averaged over sectors of the boundary. Every gap with $\Sigma C=-1$ holds a channel around the whole flake, and all channels circulate in the same sense. Of the gaps with $\Sigma C=0$, the one below the twin at $t'=t$ holds no in-gap states (g), and the one above it at $t'=2t$ holds 45 edge states, none of which encircles the flake (f, one representative state).}
\end{figure}

\emph{Flake.} On a flake of 613 cells, we compute the local Chern marker~\cite{Bianco2011}
\begin{equation}
C(\mathbf R)=-\frac{2\pi i}{A_c}\sum_{\mathbf r\in\mathbf R}\langle\mathbf r|[PXP,PYP]|\mathbf r\rangle,
\end{equation}
with $\mathbf R$ a unit cell of area $A_c$, $P$ the projector onto the states between the two gaps around the even twin, and the overall sign fixed by the bulk Chern numbers above.
Deep inside the flake it gives $-1.00$, $0.00$, and $+1.00$, the reversal of the twin (Fig.~\ref{sm:fig:Sedge_flake}).
We call an in-gap state encircling if at least 11 of 12 angular sectors of the flake each hold more than $2\%$ of its weight.
In every gap with a nonzero Chern sum all in-gap states encircle the flake and circulate in the same sense.
In the gap with zero Chern sum at $t'=2t$ the 45 in-gap states cover at most two thirds of the boundary, and none encircles it.
Of these states, 40 remain without encircling the flake when dangling triangles are removed from the boundary, so they belong to the dimer-dominated edge, and they do not form a channel.

\begin{figure}[h]
\includegraphics[width=\textwidth]{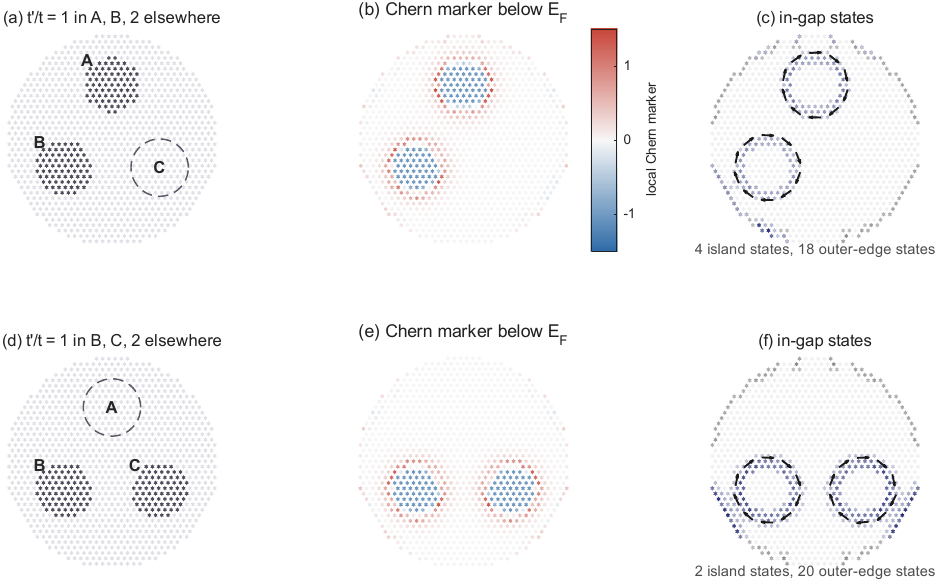}
\caption{\label{sm:fig:Sedge_switch}Switching channels with a real coupling. One connected sample of 1047 cells with the same loop current everywhere and $t'=2t$ in the background. (a,d)~Two settings of the three islands A, B, and C, with $t'=t$ in the dark islands and the islands kept at $2t$ dashed. (b,e)~Local Chern marker of all states below $E_F=-0.03t$, which is $-1$ inside the dark islands and $0$ inside the dashed one. (c,f)~In-gap states for $-0.11t<E<0.05t$, split into states on the rims of the islands (dark shading, with their current averaged over sectors of the rim, arrows) and states on the outer boundary (gray). Each island set to $t'=t$ is ringed by a one-way channel, all channels circulate in the same sense, and the outer boundary holds only edge states without a channel. Where an island comes close to the outer boundary, its channel also has some weight there.}
\end{figure}

\emph{Switch array.} The gaps above the even twin at $t'=t$ and $t'=2t$ overlap in $-0.13t<E<0.07t$, with Chern sums $-1$ and $0$.
We therefore take one connected sample of 1047 cells with the same loop current everywhere, $t'=2t$ in the background, and three islands of radius $4.6$ cells in which $t'$ is set to $t$ or kept at $2t$ (Fig.~\ref{sm:fig:Sedge_switch}).
The local Chern marker of all states below $E_F=-0.03t$ is $-1.00$ inside every island set to $t'=t$ and $0.00$ inside the island kept at $2t$.
In-gap states circulate around the rim of each island set to $t'=t$ and carry no weight on the rim of the others, and changing which islands are set moves the channels with them.
The direction of every channel is fixed by the chiral term, so at a given energy all channels circulate in the same sense.

The codes used for the numerical results are available from the authors upon request.

\end{document}